\documentclass[10pt]{article}
\usepackage{graphicx}
\def\la{\mathrel{\mathchoice {\vcenter{\offinterlineskip\halign{\hfil%
$\displaystyle##$\hfil\cr<\cr\sim\cr}}}{\vcenter{\offinterlineskip%
\halign{\hfil$\textstyle##$\hfil\cr<\cr\sim\cr}}}{\vcenter{%
\offinterlineskip\halign{\hfil$\scriptstyle##$\hfil\cr<\cr\sim\cr}}}%
{\vcenter{\offinterlineskip\halign{\hfil$\scriptscriptstyle##$\hfil\cr%
<\cr\sim\cr}}}}}
\def\ga{\mathrel{\mathchoice {\vcenter{\offinterlineskip\halign{\hfil%
$\displaystyle##$\hfil\cr>\cr\sim\cr}}}{\vcenter{\offinterlineskip%
\halign{\hfil$\textstyle##$\hfil\cr>\cr\sim\cr}}}{\vcenter{%
\offinterlineskip\halign{\hfil$\scriptstyle##$\hfil\cr>\cr\sim\cr}}}%
{\vcenter{\offinterlineskip\halign{\hfil$\scriptscriptstyle##$\hfil\cr%
>\cr\sim\cr}}}}}

\begin{document}
\def\baselinestretch{1.2}\large\normalsize

\begin{flushright}Accepted in Icarus, November 4, 2006\end{flushright}

\begin{center}

{\bf Shock-Wave Heating Model for Chondrule Formation: \\Hydrodynamic Simulation of Molten Droplets exposed to Gas Flows}

\vspace{1cm}

Hitoshi Miura,$^{\rm a,1,*}$ Taishi Nakamoto$^{\rm b,c}$

\vspace{1cm}

\small

${}^{\rm a}$Department of Physics, Kyoto University, Kitashirakawa, Sakyo, Kyoto 606-8502, Japan

${}^*$Corresponding Author E-mail address: miurah@tap.scphys.kyoto-u.ac.jp

${}^{\rm b}$Center for Computational Sciences, University of Tsukuba, Tsukuba, Ibaraki 305-8577, Japan

${}^{\rm c}$Present Address: Department of Earth and Planetary Sciences, Tokyo Institute of Technology, Meguro, Tokyo 152-8551, Japan

\vspace{1cm}

\end{center}

${}^1$Research Fellow of the Japan Society for the Promotion of Science

\vspace{1cm}

Pages: 50

Tables: 1

Figures: 13

\normalsize
\clearpage

\noindent {\bf Proposed Running Head:} Hydrodynamics of Molten Droplets in Gas Flow

\vspace{1cm}

\noindent {\bf Editorial correspondence to:}

\noindent Dr. Hitoshi Miura

\noindent Theoretical Astrophysics Group, Department of Physics, Kyoto University

\noindent Kitashirakawa, Sakyo, Kyoto 606-8502, Japan

\noindent Phone: +81-75-753-3880

\noindent Fax: +81-75-753-3886

\noindent E-mail: miurah@tap.scphys.kyoto-u.ac.jp

\clearpage

\abstract{
Millimeter-sized, spherical silicate grains abundant in chondritic meteorites, which are called as chondrules, are considered to be a strong evidence of the melting event of the dust particles in the protoplanetary disk. One of the most plausible scenarios is that the chondrule precursor dust particles are heated and melt in the high-velocity gas flow (shock-wave heating model). We developed the non-linear, time-dependent, and three-dimensional hydrodynamic simulation code for analyzing the dynamics of molten droplets exposed to the gas flow. We confirmed that our simulation results showed a good agreement in a linear regime with the linear solution analytically derived by Sekiya et al. (2003). We found that the non-linear terms in the hydrodynamical equations neglected by Sekiya et al. (2003) can cause the cavitation by producing negative pressure in the droplets. We discussed that the fragmentation through the cavitation is a new mechanism to determine the upper limit of chondrule sizes. We also succeeded to reproduce the fragmentation of droplets when the gas ram pressure is stronger than the effect of the surface tension. Finally, we compared the deformation of droplets in the shock-wave heating with the measured data of chondrules and suggested the importance of other effects to deform droplets, for example, the rotation of droplets. We believe that our new code is a very powerful tool to investigate the hydrodynamics of molten droplets in the framework of the shock-wave heating model and has many potentials to be applied to various problems. \\

}

\clearpage

\section{Introduction}
Chondrules are millimeter-sized, once-molten, spherical-shaped grains mainly composed of silicate material. They are abundant in chondritic meteorites, which are the majority of meteorites falling onto the Earth. They are considered to have formed from chondrule precursor dust particles about $4.56 \times 10^9 \, {\rm yr}$ ago in the solar nebula (Amelin et al. 2002); they were heated and melted through flash heating events in the solar nebula and cooled again to solidify in a short period of time (e.g., Jones et al. 2000 and references therein). So they must have great information on the early history of our solar system. Since it is naturally expected that protoplanetary disks around young stars in star forming regions have similar dust particles and processes, the study of chondrule formation may provide us much information on the planetary system formation. Chondrules have many features: physical properties (sizes, shapes, densities, degree of magnetization, etc.), isotopic compositions (oxygen, nitrogen, rare gases, etc.), mineralogical and petrologic features (structures, crystals, degrees of alteration, relicts, etc.), and so forth, each of which should be a clue that helps us to reveal their own formation process and that of the planetary system. To reveal their formation history, many works have been carried out observationally, experimentally, and theoretically. 


Shock-wave heating model is considered to be one of the most plausible models for chondrule formation (Boss 1996, Jones et al. 2000). This model has been investigated theoretically by many authors (Hood \& Horanyi 1991, 1993, Ruzmaikina \& Ip 1994, Tanaka et al. 1998, Hood 1998, Iida et al. 2001, Desch \& Connolly 2002, Miura et al. 2002, Ciesla \& Hood 2002, Miura \& Nakamoto 2005, 2006). One of the special features of this model is that chondrule precursor dust particles are exposed to the high-velocity rarefied gas flow. In the gas flow, the gas frictional heating takes place to heat and melt the precursor dust particles (the condition to melt silicate dust particles was derived by Iida et al. 2001). Therefore, it is naturally thought that the ram pressure of the gas flow affects the molten dust particles. Susa \& Nakamoto (2002) estimated the maximum radius of the molten droplet above which the droplet should be destroyed by the ram pressure. They analyzed the non-linear phenomena like a fragmentation by an order of magnitude estimation. On the contrary, Sekiya et al. (2003) derived a linear solution of the deformation and the internal flow of the molten droplet exposed to the rarefied gas flow. They analytically solved the hydrodynamical equations assuming that the non-linear terms of the hydrodynamical equations as well as the surface deformation are sufficiently small so that linearized equations are appropriate. 

The hydrodynamics of the molten droplet exposed to the rarefied gas flow is an important aspect of the shock-wave heating model. The internal flow causes the homogenization of the chemical/isotopic compositions and temperature distribution in the molten precursor dust particles. The deformation of the molten droplet might result into the deformed chondrules. Moreover, the fragmentation of the molten droplet gives an upper limit of chondrules as pointed by Susa \& Nakamoto (2002). However, the analysis of Susa \& Nakamoto (2002) did not take into account the hydrodynamics of the droplet and the formulation of Sekiya et al. (2003) cannot be applied to such non-linear phenomena like the fragmentation. Additionally, the effects of non-linear terms in the hydrodynamical equations, which was neglected in Sekiya et al. (2003), might be important in the hydrodynamics. In order to solve the hydrodynamical equations including the non-linear terms, the numerical simulations can be a powerful tool. The purpose of this study is to develop a numerical model of the molten droplet exposed to the high-velocity rarefied gas flow. 

We describe the physical model in \S 2 and the basic equations in \S 3. The numerical model is summarized in \S 4. We show the calculation results in \S 5 and some discussions in \S 6. Finally, we make conclusions in \S 7. Some numerical techniques we use in our model are summarized in appendixes.

\section{Physical Model}
\label{sec:physical_model}
We are considering the molten silicate dust particles exposed to the high-velocity rarefied gas flow. These dust particles are, of course, initially solid and then melted due to the gas frictional heating. In this paper, we do not consider how the solid particles melt in the gas flow. We assume the completely molten droplets (no solid lumps inside) and the physical properties of the droplet (the coefficient of viscosity and the fluid surface tension coefficient) are constant. The droplet behaves as an incompressible fluid because the sound speed ($\sim$ a few ${\rm km \, s^{-1}}$, Murase \& McBirney 1973) is much larger than the fluid velocity expected in the droplet ($\sim$ a few tens ${\rm cm \, s^{-1}}$, Sekiya et al. 2003). 

The ram pressure of the gas flow is acting on the droplet surface exposed to the high-velocity gas flow. It should be noted that the gas flow we consider here does not follow the hydrodynamical equations in the spatial scale of chondrules because the nebular gas is too rarefied. The mean free path of the nebula gas can be estimated by $l = 1 / (ns)$, where $s$ is the collisional cross section of gas molecules and $n$ is the number density of the nebular gas. In the minimum mass solar nebula model (Hayashi et al. 1985), the number density of the nebula gas at 1 AU from the central star is about $n \simeq 10^{14} \, {\rm cm^{-3}}$. The collisional cross section of the hydrogen molecule is roughly estimated as $s \simeq 10^{-16} \, {\rm cm^{2}}$ (e.g., Hollenbach \& McKee 1979). So, we obtain $l \simeq 100 \, {\rm cm}$. On the contrary, the typical size of chondrules is about a few $100 \, {\rm \mu m}$ (e.g., Rubin \& Keil 1984). Namely, the object which disturbs the gas flow is much smaller than the mean free path of the gas. This condition corresponds to a free molecular flow in which gas molecules scattered and reemitted from the droplet surface do not disturb the free stream velocity distribution. Therefore, in our model, the ram pressure acting on the droplet surface per unit area is explicitly given by the momentum flux of the molecular gas flow $p_{\rm fm} = \rho_{\rm g} v_{\rm rel}^2$, where $\rho_{\rm g}$ is the nebula gas density and $v_{\rm rel}$ is the relative velocity between the gas and the mass center of the droplet. This force causes the deceleration of the center of mass. In the coordinate system co-moving with the center of mass (we adopt this coordinate in our model), the apparent gravitational acceleration should be considered. 

Strictly speaking, the relative velocity of a gas molecule and a surface fluid element is not $v_{\rm rel}$, since the latter has a non-zero velocity in the rest frame of the mass center of the droplet. However, the liquid velocity is much smaller than the gas molecular velocity, and is negligible (Sekiya et al. 2003). In hypersonic free molecular flow, the thermal and reflected velocities of gas molecules are also sufficiently small and are ignored.

\section{Basic Equations\label{sec:basic_equation}}

\subsection{Equation of Continuity}
\label{sec:mass_conservation}
Generally, Eulerian methods like the CIP scheme (e.g., Yabe et al. 2001) use a color function $\phi$ in the multi-phase analysis. For example, when a fluid of phase $k$ occupies a certain region, the color function takes $\phi_k = 1$ inside the region and $\phi_k = 0$ outside. In this paper, we define $\phi=1$ for inside the molten dust particle and $\phi=0$ for outside (ambient region). Using the color function, the density of the fluid element $\rho$ is given by
\begin{equation}
\rho = \rho_{\rm d} \times \phi + \rho_{\rm a} \times ( 1 - \phi ), 
\label{eq:density}
\end{equation}
where the subscripts ``d" and ``a" indicate the inherent values for the molten silicate dust particle (droplet) and the ambient region, respectively. The other physical values of the fluid element (the coefficient of viscosity $\mu$ and the sound speed $c_{\rm s}$) are given in the same manner, namely, $\mu = \mu_{\rm d} \times \phi + \mu_{\rm a} \times ( 1 - \phi )$ and $c_{\rm s} = c_{\rm s,d} \times \phi + c_{\rm s,a} \times ( 1 - \phi )$, respectively\footnote{In the numerical simulation, an undesirable situation that the hydrostatic pressure $p$ becomes much larger than the value of the ambient region in spite of $\phi \ll 1$ could occur. When it occurred, we set the sound speed of the fluid element $c_{\rm s} \rightarrow c_{\rm s,d}$ in order to avoid the difficulty in the numerical simulation.}.

We need an equation that describes the time evolution of the color function $\phi$. Substituting Eq. (\ref{eq:density}) to the equation of continuity, 
\begin{equation}
\frac{ \partial \rho }{ \partial t } + \mbox{\boldmath $\nabla$} \cdot ( \rho \mbox{\boldmath $u$}) = 0, 
\label{eq:mass_conservation}
\end{equation}
where $\mbox{\boldmath $u$}$ is the velocity, we obtain 
\begin{equation}
\frac{ \partial \phi }{ \partial t } + \mbox{\boldmath $\nabla$} \cdot ( \phi \mbox{\boldmath $u$} ) 
=
- \frac{ \rho_{\rm a} }{ \rho_{\rm d} - \rho_{\rm a} } \mbox{\boldmath $\nabla$} \cdot \mbox{\boldmath $u$}.
\label{eq:mass_conservation_phi}
\end{equation}
Integrating Eq. (\ref{eq:mass_conservation_phi}) over the closed region with the boundary condition of $\mbox{\boldmath $u$} = \mbox{\boldmath $0$}$, we obtain
\begin{equation}
\frac{ \partial }{ \partial t } \int \phi d\mbox{\boldmath $r$} = 0.
\label{eq:mass_conservation_phi_integrate}
\end{equation}
Therefore, it is found that the total amount of the color function $\phi$ in the closed region is conserved. Then, we rewrite Eq. (\ref{eq:mass_conservation_phi}) as
\begin{equation}
\frac{ \partial \phi }{ \partial t } + ( \mbox{\boldmath $u$} \cdot \mbox{\boldmath $\nabla$} ) \phi
=
- \bigg(1 + \frac{ \rho_{\rm a} }{ \rho_{\rm d} - \rho_{\rm a} } \frac{ 1 }{ \phi } \bigg) \phi
\mbox{\boldmath $\nabla$} \cdot \mbox{\boldmath $u$}.
\label{eq:mass_conservation_phi2}
\end{equation}
The typical density of the silicate melt is $\rho_{\rm d} = 3 \, {\rm g \, cm^{-3}}$, while the density of the ambient region is extremely low (e.g., typical nebular gas density is about $10^{-9} \, {\rm g \, cm^{-3}}$). Therefore, we obtain $\rho_{\rm a} / ( \rho_{\rm d} - \rho_{\rm a} ) \ll 1$. Additionally, we can consider that the value of the color function $\phi$ is order of unity because we especially focus on the hydrodynamics of the molten silicate dust particle in this study. Therefore, the second term on the right hand side of Eq. (\ref{eq:mass_conservation_phi2}) can be ignored. Finally, we obtain the different form of the equation of continuity as
\begin{equation}
\frac{ \partial \phi }{ \partial t } + ( \mbox{\boldmath $u$} \cdot \mbox{\boldmath $\nabla$} ) \phi
=
- \phi \mbox{\boldmath $\nabla$} \cdot \mbox{\boldmath $u$}.
\label{eq:mass_conservation_phi3}
\end{equation}
It should be noted that the total amount of $\phi$ is conserved even in the approximated form. Eq. (\ref{eq:mass_conservation_phi3}) gives the time evolution of the color function $\phi$. After calculating $\phi$, we can obtain the density $\rho$ by Eq. (\ref{eq:density}).

\subsection{Equation of Motion}
The local velocity of the fluid element is changed by the pressure gradient, the viscous force, the surface tension, and the ram pressure of the high-velocity molecular gas flow. The ram pressure exerted on the surface of the droplet is given by (Sekiya et al. 2003)
\begin{equation}
\mbox{\boldmath $F_{\rm g}$} = -p_{\rm fm} ( \mbox{\boldmath $n_{\rm i}$} \cdot \mbox{\boldmath $n_{\rm g}$} ) \mbox{\boldmath $n_{\rm g}$} \delta(\mbox{\boldmath $r$} - \mbox{\boldmath $r_{\rm i}$} )
~~~ {\rm for} ~ \mbox{\boldmath $n_{\rm i}$} \cdot \mbox{\boldmath $n_{\rm g}$} \le 0, 
\label{eq:gas_ram_pressure}
\end{equation}
where $\mbox{\boldmath $n_{\rm i}$}$ is the unit normal vector of the surface of the droplet, $\mbox{\boldmath $n_{\rm g}$}$ is the unit vector pointing the direction in which the gas flows, and $\mbox{\boldmath $r_{\rm i}$}$ is the position of the liquid-gas interface. The delta function $\delta ( \mbox{\boldmath $r$} - \mbox{\boldmath $r_{\rm i}$} )$ means that the ram pressure works only at the interface. The ram pressure does not work for $\mbox{\boldmath $n_{\rm i}$} \cdot \mbox{\boldmath $n_{\rm g}$} > 0$ because it indicates the opposite surface which does not face the molecular flow. The ram pressure causes the deceleration of the center of mass of the droplet. In our coordinate system co-moving with the center of mass, the apparent gravitational acceleration $\mbox{\boldmath $g$}$ should appear in the equation of motion. 

The surface tension is given as (e.g., Brackbill et al. 1992)
\begin{equation}
\mbox{\boldmath $F_{\rm s}$} = - \gamma \kappa \mbox{\boldmath $n_{\rm i}$} 
\delta(\mbox{\boldmath $r$} - \mbox{\boldmath $r_{\rm i}$} ),
\label{eq:surface_tension}
\end{equation}
where $\gamma$ is the fluid surface tension coefficient and $\kappa$ is the local surface curvature. 

Finally, the equation of motion is written in a form
\begin{equation}
\frac{ \partial \mbox{\boldmath $u$} }{ \partial t } + ( \mbox{\boldmath $u$} \cdot \mbox{\boldmath $\nabla$} ) \mbox{\boldmath $u$} = \frac{ - \mbox{\boldmath $\nabla$} p + \mu \Delta \mbox{\boldmath $u$} + \mbox{\boldmath $F_{\rm s}$} + \mbox{\boldmath $F_{\rm g}$} }{ \rho } + \mbox{\boldmath $g$}, 
\label{eq:equation_of_motion}
\end{equation}
where $p$ is the pressure and $\mu$ is the coefficient of viscosity. In deriving the viscous term $\mu \Delta \mbox{\boldmath $u$}$, we assume that the viscous coefficient $\mu$ do not change noticeably and the contribution of $\mbox{\boldmath $\nabla$} \cdot \mbox{\boldmath $u$}$ in the viscosity is negligibly small in the droplet (Landau \& Lifshitz 1987).

\subsection{Equation of State}
We can obtain the equation that describes  the time evolution of the pressure $p$ from the equation of state, which is given by
\begin{equation}
\frac{ dp }{ d\rho } = c_{\rm s}^2, 
\label{eq:equation_of_state}
\end{equation}
where $c_{\rm s}$ is the sound speed. We can rewrite it into
\begin{equation}
\frac{ dp }{ dt } = c_{\rm s}^2 \frac{ d\rho }{ dt }, 
\label{eq:equation_of_motion2}
\end{equation}
where $d/dt = \partial / \partial t + (\mbox{\boldmath $u$} \cdot \mbox{\boldmath $\nabla$} )$. Substituting the mass conservation equation (Eq. \ref{eq:mass_conservation}), we obtain 
\begin{equation}
\frac{ \partial p }{ \partial t } + ( \mbox{\boldmath $u$} \cdot \mbox{\boldmath $\nabla$} ) p 
= - \rho c_{\rm s}^2 \mbox{\boldmath $\nabla$} \cdot \mbox{\boldmath $u$}. 
\label{eq:equation_of_pressure}
\end{equation}

\section{Numerical Model}

\subsection{Droplet and Gas Flow}
\label{sec:droplet_gasflow}
\begin{figure}[]
\center
\includegraphics[scale=.8, angle=0.]{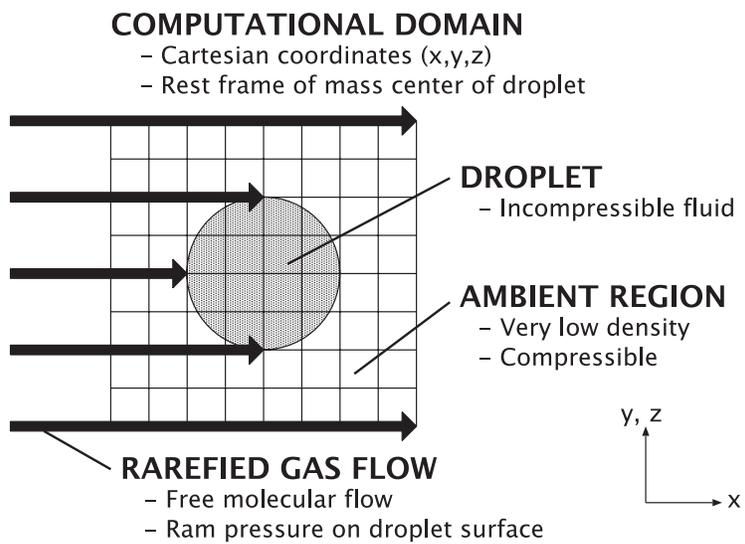}
\caption{Schematic picture of our numerical model and the coordinate system.}
\label{fig:numerical_model}
\end{figure}

We consider the molten silicate dust particle exposed to the high-velocity rarefied gas flow. As we mentioned in \S \ref{sec:physical_model}, the gas flow is the free molecular flow because the mean free path of the gas molecules is larger than the typical particle size. The gas flow comes from the upstream and terminates when it collide with the dust surface. Figure \ref{fig:numerical_model} shows the schematic picture of our numerical model and the coordinate system. The black thick arrows show the stream line of the gas flow. The gas flow gives the momentum to the dust particle at the area where it terminates. Therefore, the ram pressure exerted on the dust surface $\mbox{\boldmath $F_{\rm g}$}$ can be calculated easily, even if we do not solve the hydrodynamic equations for the ambient region. The numerical treatment of the ram pressure is described in \S \ref{appendix:gas_drag_force}. 

Of course, we know that the gas flow has not only the free molecular motion (see Fig. \ref{fig:numerical_model}), but also the random velocity caused by the thermal motion. It causes the hydrostatic pressure at the dust surface. However, the hydrostatic pressure of the ambient region is so small that it can be disregarded to the gas ram pressure. The typical value for the hydrostatic pressure in the solar nebula is about $p_{\rm SN} = n k_{\rm B} T \simeq 4 \, {\rm dyne \, cm^{-2}}$ for $n = 10^{14} \, {\rm cm^{-3}}$ and $T = 300 \, {\rm K}$, where $n$ is the number density of the nebula gas and $T$ is the gas temperature. On the contrary, the ram pressure of the gas flow is about $p_{\rm fm} \sim 4000 \, {\rm dyne \, cm^{-2}}$ for the shock-wave heating model for chondrule formation (e.g., Uesugi et al. 2003). Additionally, the hydrostatic pressure of the nebula gas is also much smaller than that inside the molten droplet. We have the well known equation for the hydrostatic pressure inside the droplet as $p_0 = 2 \gamma / r_0$, where $r_0$ is the droplet radius. Substituting $\gamma = 400 \, {\rm dyne \, cm^{-1}}$ for the molten silicates (Murase \& McBirney 1973) and $r_0 = 500 \, {\rm \mu m}$ for the typical chondrule radius, we obtain $p_0 = 1.6 \times 10^4 \, {\rm dyne \, cm^{-2}}$. Therefore, the hydrostatic pressure of the nebula gas is much smaller than that inside the molten droplet and the ram pressure of the gas flow. Therefore, we consider that the effect of $p_{\rm SN}$ is negligibly small for the hydrodynamics of the molten droplet and neglect that. 

\hspace{1cm}{\bf [Figure \ref{fig:numerical_model}]}

\subsection{Ambient Region}
\label{sec:vacuum}
As we mentioned in the previous subsection, what we should do for investigating the dynamics of the molten droplet exposed to the free molecular flow is just to solve the hydrodynamical equations for the molten droplet. Namely, there is no need to solve the hydrodynamical equations for the ambient region because the gas ram pressure exerted at the droplet surface can be calculated without doing that. 

In that case, how can we treat the outside of the droplet? In our numerical model, we neglect the influence of the ambient region on the dynamics of the molten droplet (see \S \ref{sec:droplet_gasflow}). The best way to do that is to place nothing in the ambient region. It means that the fluid density outside the droplet is zero (see Eq. \ref{eq:density}). However, the equation of motion cannot be solved if $\rho = 0$ (see Eq. \ref{eq:equation_of_motion}). Therefore, we must put the density of the ambient region $\rho_{\rm a} > 0$ in order to solve the hydrodynamical equations without any other special techniques. At the same time, $\rho_{\rm a}$ should be small enough not to affect the dynamics of the molten droplet. 

Additionally, we should pay attention to the sound speed of the ambient region, $c_{\rm s,a}$. If $c_{\rm s,a}$ is large enough for the ambient region to behave as an incompressible fluid, the influences of the boundaries of the computational domain would travel to the molten droplet in an instant. We cannot say that this situation correctly simulates the molten droplet in the shock-wave heating. On the contrary, if $c_{\rm s,a}$ is small enough to compress (or expand) the ambient region without any restitution, the dynamics of the droplet are not affected by the boundaries. Therefore, in our numerical model, $c_{\rm s,a}$ should be small. 

We put $\rho_{\rm a} = 10^{-6} \, {\rm g \, cm^{-3}}$ and $c_{\rm s,a} = 10^{-5} \, {\rm cm \, s^{-1}}$ as standard values for the numerical simulations (see \S \ref{appendix:model_parameter}).

\subsection{Coordinate system}
\label{sec:coordinate_system}
We set the molten droplet at the center of a cubic computational domain (see Fig. \ref{fig:numerical_model}). We adopt the Cartesian coordinate system ($x,y,z$), which is co-moving with the mass center of the droplet. Since the ram pressure of the gas flow explicitly appears in the equation of motion (Eq. \ref{eq:equation_of_motion}), the ambient region ($\phi=0$) does not play any important role to the droplet. 

In our model, since we examine the hydrodynamics of the droplet, it is no use to calculate outside of the droplet. It indicates that we consume some amount of the computational time to calculate the ambient region which is not important in our purpose. However, above model allows us to adopt the unified numerical procedure for compressible and incompressible fluid, which has been developed based on the concept of the CIP scheme (Yabe \& Wang 1991). Therefore, we choose to sacrifice some computational time in order to make the coding process simple.

\subsection{Scheme}

In the numerical scheme, we especially should pay attention to following points. First, the numerical scheme for the equation of continuity should guarantee the mass conservation. If not, the droplet radius might be changed by some numerical effects. Second, the discontinuity in the profile of $\phi$ between the droplet and the ambient region should be kept as the sharp profile. If some numerical effects make the interface diffuse, it would be difficult to introduce the surface tension. Final point is the incompressiblity of the droplet. 

The equation of continuity in the form of Eq. (\ref{eq:mass_conservation_phi3}) can be rewritten as
\begin{equation}
\frac{ \partial \phi }{ \partial t } + \mbox{\boldmath $\nabla$} \cdot ( \phi \mbox{\boldmath $u$} ) = 0,
\label{eq:mass_conservation_phi4}
\end{equation}
where the total amount of $\phi$ in a closed region should be conserved. The R-CIP-CSL2 scheme is one of the recent versions of the CIP scheme, which guarantees the exact conservation even in the framework of a semi-Lagrangian scheme (Nakamura et al. 2001). In order to solve Eq. (\ref{eq:mass_conservation_phi4}), we adopt the R-CIP-CSL2 scheme and briefly describe this scheme in appendix \ref{appendix:r-cip_scheme}. Additionally, we make a proper correction for the velocity $\mbox{\boldmath $u$}$ before solving Eq. (\ref{eq:mass_conservation_phi4}) in order to keep the incompressibility of the droplet (see appendix \ref{appendix:divergence_free}). 

The other two hydrodynamical equations can be separated into the advection phase and the non-advection phase. The advection phases of the equation of motion (Eq. \ref{eq:equation_of_motion}) and the equation of state (Eq. \ref{eq:equation_of_pressure}) are written as
\begin{equation}
\frac{ \partial \mbox{\boldmath $u$} }{ \partial t } + ( \mbox{\boldmath $u$} \cdot \mbox{\boldmath $\nabla$} ) \mbox{\boldmath $u$} = 0 , 
~~~
\frac{ \partial p }{ \partial t } + ( \mbox{\boldmath $u$} \cdot \mbox{\boldmath $\nabla$} ) p 
= 0 .
\label{eq:advection_phase}
\end{equation}
We solve above equations using the R-CIP scheme, which is the oscillation preventing scheme for advection equation (Xiao et al. 1996b, see appendix \ref{appendix:cip_scheme}). The non-advection phases can be written as
\begin{equation}
\frac{ \partial \mbox{\boldmath $u$} }{ \partial t }
= - \frac{ \mbox{\boldmath $\nabla$} p }{ \rho } + \frac{ \mbox{\boldmath $Q$}}{ \rho } ,
~~~
\frac{ \partial p }{ \partial t } 
= - \rho c_{\rm s}^2 \mbox{\boldmath $\nabla$} \cdot \mbox{\boldmath $u$} ,
\label{eq:non-advection_phase}
\end{equation}
where $\mbox{\boldmath $Q$}$ is the summation of forces except for the pressure gradient. The problem intrinsic in incompressible fluid is in the high sound speed in the pressure equation. Yabe \& Wang (1991) introduced an excellent approach to avoid the problem (see appendix \ref{appendix:c-cup_scheme}). It is called as the C-CUP scheme (Yabe et al. 2001). 

Additionally, we introduce the anti-diffusion technique when calculating the equation of continuity by the R-CIP-CSL2 scheme. The original R-CIP-CSL2 scheme provides an excellent solution for the conservative equation (e.g., Nakamura et al. 2001). However, if the initial profile of $\phi$ has a sharp discontinuity (e.g., rectangle wave), the profile is losing its sharpness as the time step progresses. It is a result of the numerical diffusion. In order to keep the discontinuity of the profile, the anti-diffusion modification is an useful technique (e.g., Xiao \& Ikebata 2003). In this paper, we explicitly add the diffusion term with a negative diffusion coefficient to the CIP-CSL2 scheme (see appendix \ref{appendix:anti-diffusion}). We perform the one-dimensional conservative equation with or without the anti-diffusion technique in order to show the validity of this method (see appendix \ref{appendix:test_calculation}). 

\begin{figure}[]
\center
\includegraphics[scale=.6, angle=0.]{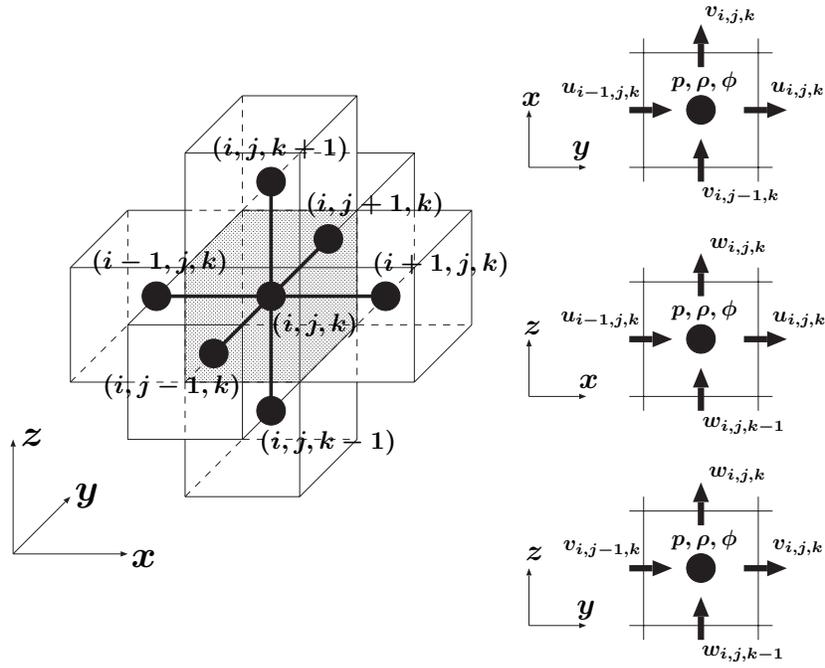}
\caption{The staggered grid adopted in our model. The scalar variables are defined at the cell center and the velocity $\mbox{\boldmath $u$} = (u, v, w)$ is defined on the cell-edge. The external force like the surface tension and the ram pressure of the gas flow are defined at the cell-center.}
\label{fig:staggered}
\end{figure}

For the multi-dimensions, we use a directional splitting technique to perform sequential one-dimensional procedure in each direction. We adopt the staggered grid for digitizing the physical variables in the hydrodynamical equations (see Fig. \ref{fig:staggered}). The scalar variables are defined at the cell-center and the velocity $\mbox{\boldmath $u$} = (u, v, w)$ is defined on the cell-edge. The surface tension and the ram pressure are defined at the cell-center. The numerical model of these two forces are shown in appendixes \ref{appendix:gas_drag_force} and \ref{appendix:surface_tension}. When calculating these two forces, the gradient field of $\phi$ is required and needs to be artificially smoothed (e.g., Yabe et al. 2001, see appendix \ref{appendix:smoothing} in this paper). Note that the smoothing is done only at the calculation of these two forces and the original profile with sharp discontinuity is unchanged. 

\hspace{1cm}{\bf [Figure \ref{fig:staggered}]}

\section{Results}

\subsection{Input Parameters and Initial Settings\label{sec:initial_parameter}}
We investigate various cases about the droplet radius, $r_0 = 100$, $200$, $500$, $1000$, $2000$, $5000 \, {\rm \mu m}$, $1\, {\rm cm}$, and $2 \, {\rm cm}$. The momentum flux of the gas flow is set as $p_{\rm fm} = 4000 \, {\rm dyne \, cm^{-2}}$, which may be realized in the shock-wave heating model for relatively high-density shock waves (Uesugi et al. 2003). The coefficient of viscosity of silicate melts $\mu_{\rm d}$ strongly depends on the temperature and the chemical composition. We adopt $\mu_{\rm d} = 1.3 \, {\rm g \, cm^{-1} \, s^{-1}}$ from the model of Uesugi et al. (2003), in which they calculated the viscosity by the formulation of Bottinga \& Weill (1972) assuming the temperature $\sim 1800 \, {\rm K}$ and the chemical composition of BO type chondrule. The surface tension $\gamma = 400 \, {\rm dyne \, cm^{-1}}$ and the sound speed of the molten droplet $c_{\rm s,d} = 2 \times 10^5 \, {\rm cm \, s^{-1}}$ are the typical value of the silicate melts (Murase \& McBirney 1973). Other input parameters are listed in Table \ref{table:input_parameters}. When we consider the fragmentation process of the droplets in the gas flow, it is useful to introduce the non-dimensional parameter $W_e \equiv p_{\rm fm} r_0 / \gamma$, which is called as the Weber number and indicates the ratio of the ram pressure of the gas flow to the surface tension of the droplet. Some experiments suggested that the droplets fragment for $W_e \sim 6 - 22$ or higher (Bronshten 1983). 

\hspace{1cm}{\bf [Table \ref{table:input_parameters}]}

We assume that the gas flow suddenly affects the initially spherical droplet (see the panel (a) in Fig. \ref{fig:gasdrag_a00500_timeevo}). The horizontal and vertical axes are the $x$- and $y$-axes normalized by the initial droplet radius $r_0$. The gas flow comes from the left side of the panel. The color contour indicates the hydrostatic pressure $p$ in the unit of ${\rm dyne \, cm^{-2}}$ and the arrows are the velocity field. We have the well known equation for the initial pressure inside the droplet as $p = 2 \gamma / r_0$ and $p=0$ for outside. The red curves indicate the contour of the color function $\phi$ after smoothing. Thick solid curve indicates $\phi = 0.5$, which means the interface between the droplet and the ambient region, and the dashed and the dotted-dashed curves are $\phi = 0.1$ and $0.9$, which are drawn in order to show the effective width of the transition region between the molten droplet and the ambient region. The initial internal velocity of the droplet is set to be all zero. The computational grid points are set to be $60\times60\times60$ for the standard calculations. The boundary conditions are $\mbox{\boldmath $u$} = \mbox{\boldmath $0$}$, $p=0$, $\rho = \rho_{\rm a}$, $\phi=0$, and $\sigma = 0$.

\subsection{Time Evolution}
\begin{figure}[]
\center
\begin{tabular}{cc}
	\vspace{3mm}
	\begin{minipage}{.35\linewidth}
		\begin{center}(a) initial\end{center}
		\vspace{-5mm}
		\includegraphics[width=\linewidth, angle=-90.]{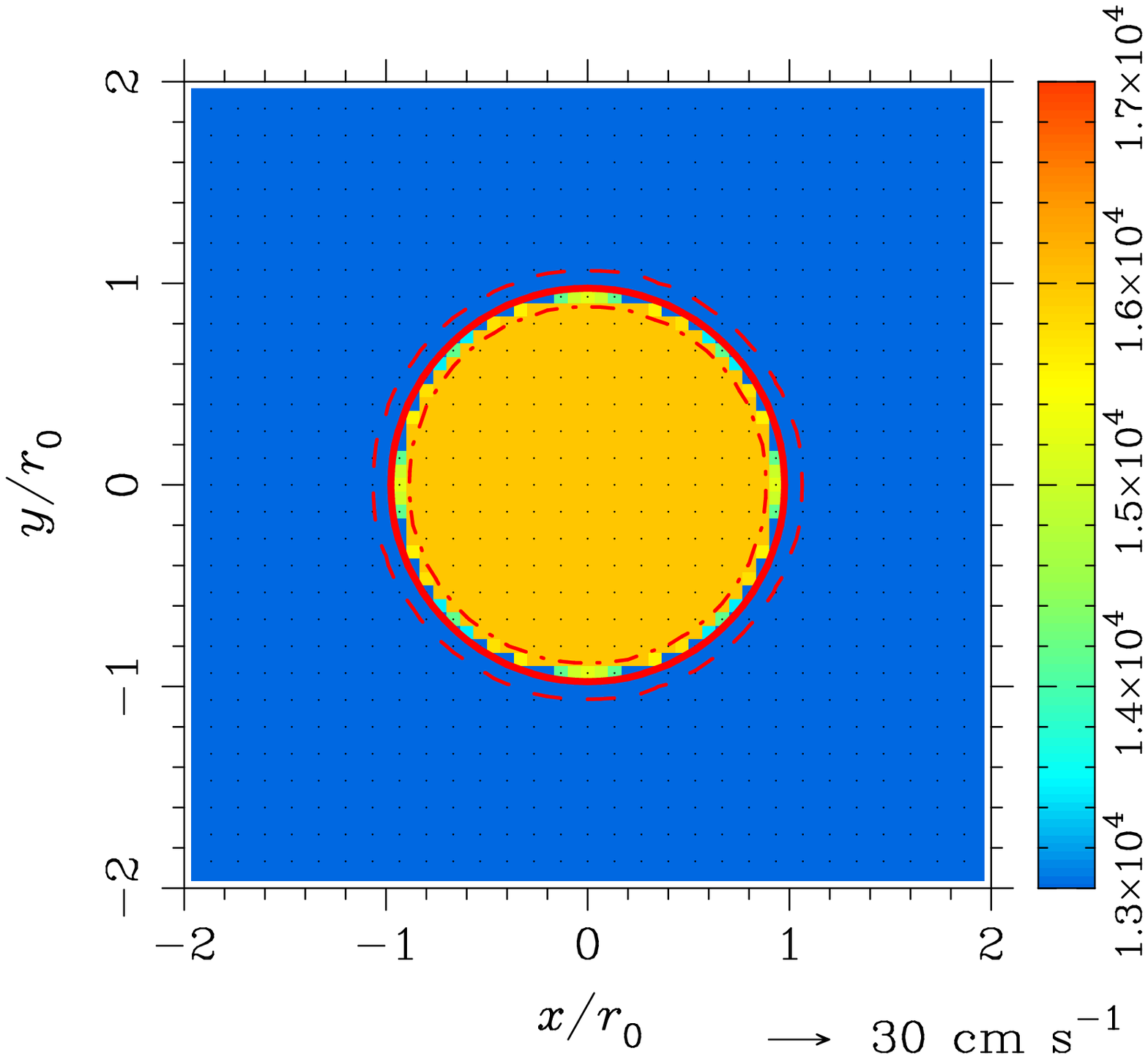}
	\end{minipage} &
	\hspace{6mm}
	\begin{minipage}{.35\linewidth}
		\begin{center}(d) 1.0 msec\end{center}
		\vspace{-5mm}
		\includegraphics[width=\linewidth, angle=-90.]{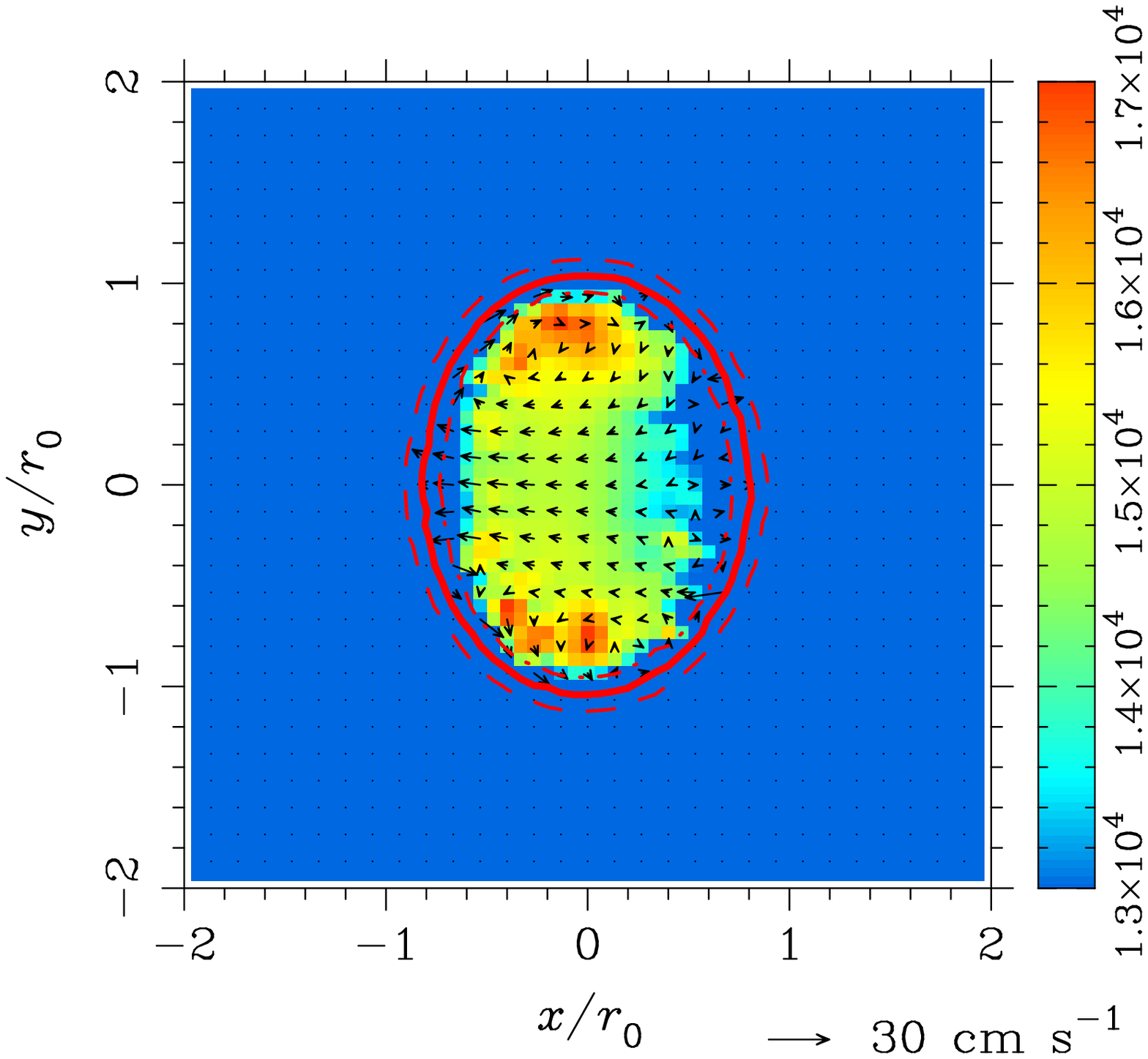}
	\end{minipage} \\
	\vspace{3mm}
	\begin{minipage}{.35\linewidth}
		\begin{center}(b) 0.3 msec\end{center}
		\vspace{-5mm}
		\includegraphics[width=\linewidth, angle=-90.]{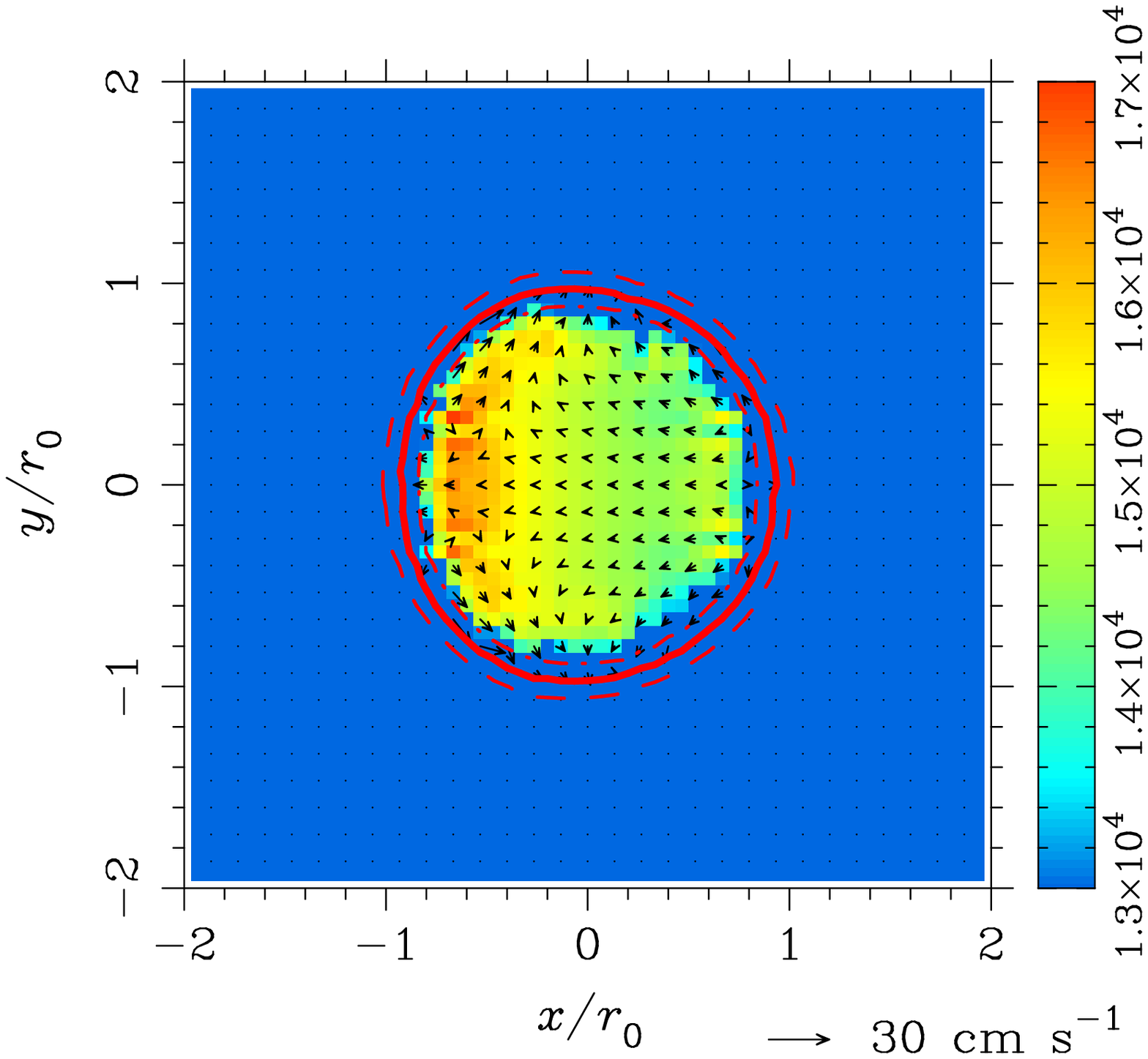}
	\end{minipage} &
	\hspace{6mm}
	\begin{minipage}{.35\linewidth}
		\begin{center}(e) 1.5 msec\end{center}
		\vspace{-5mm}
		\includegraphics[width=\linewidth, angle=-90.]{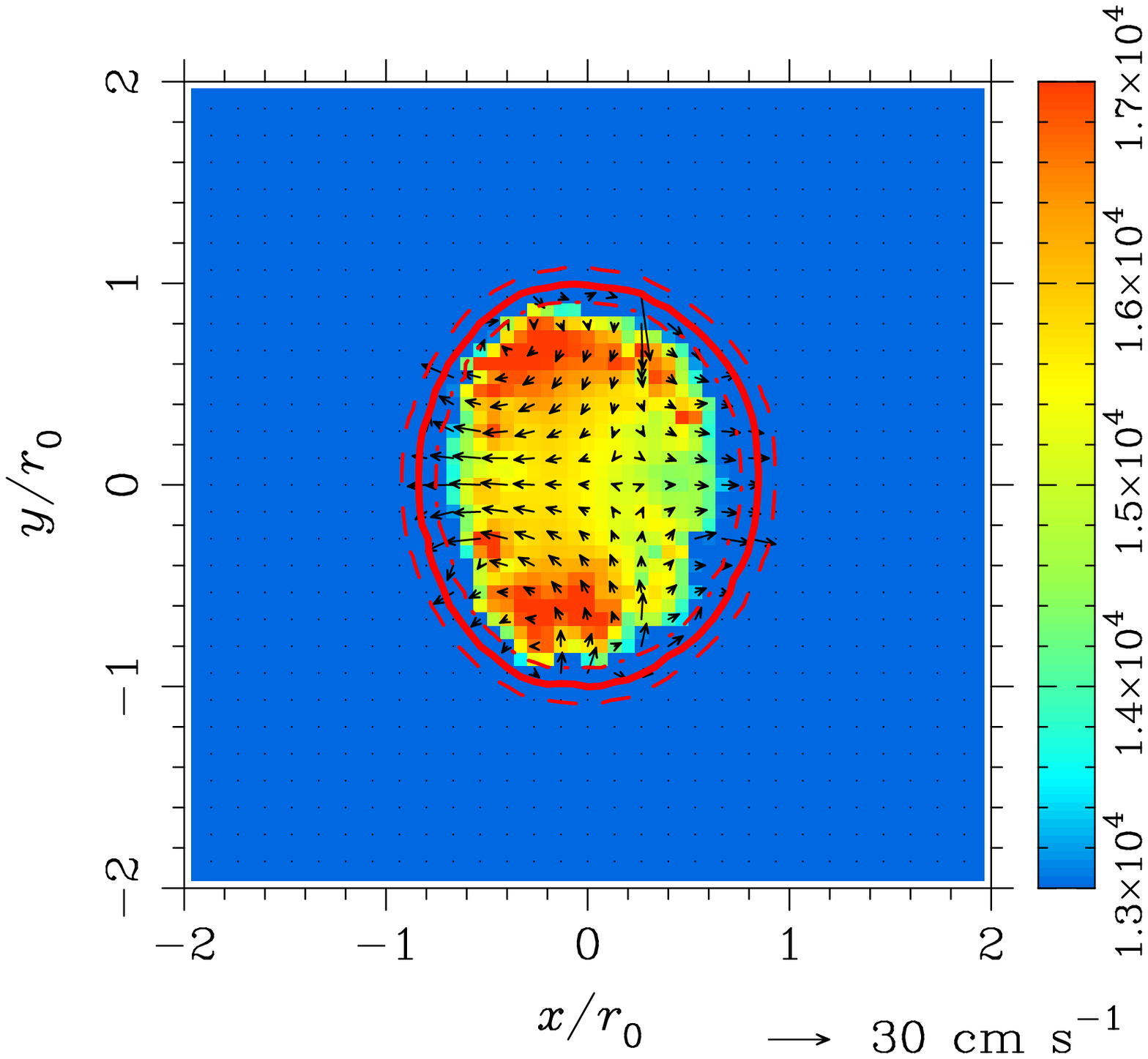}
	\end{minipage} \\
	\vspace{3mm}
	\begin{minipage}{.35\linewidth}
		\begin{center}(c) 0.6 msec\end{center}
		\vspace{-5mm}
		\includegraphics[width=\linewidth, angle=-90.]{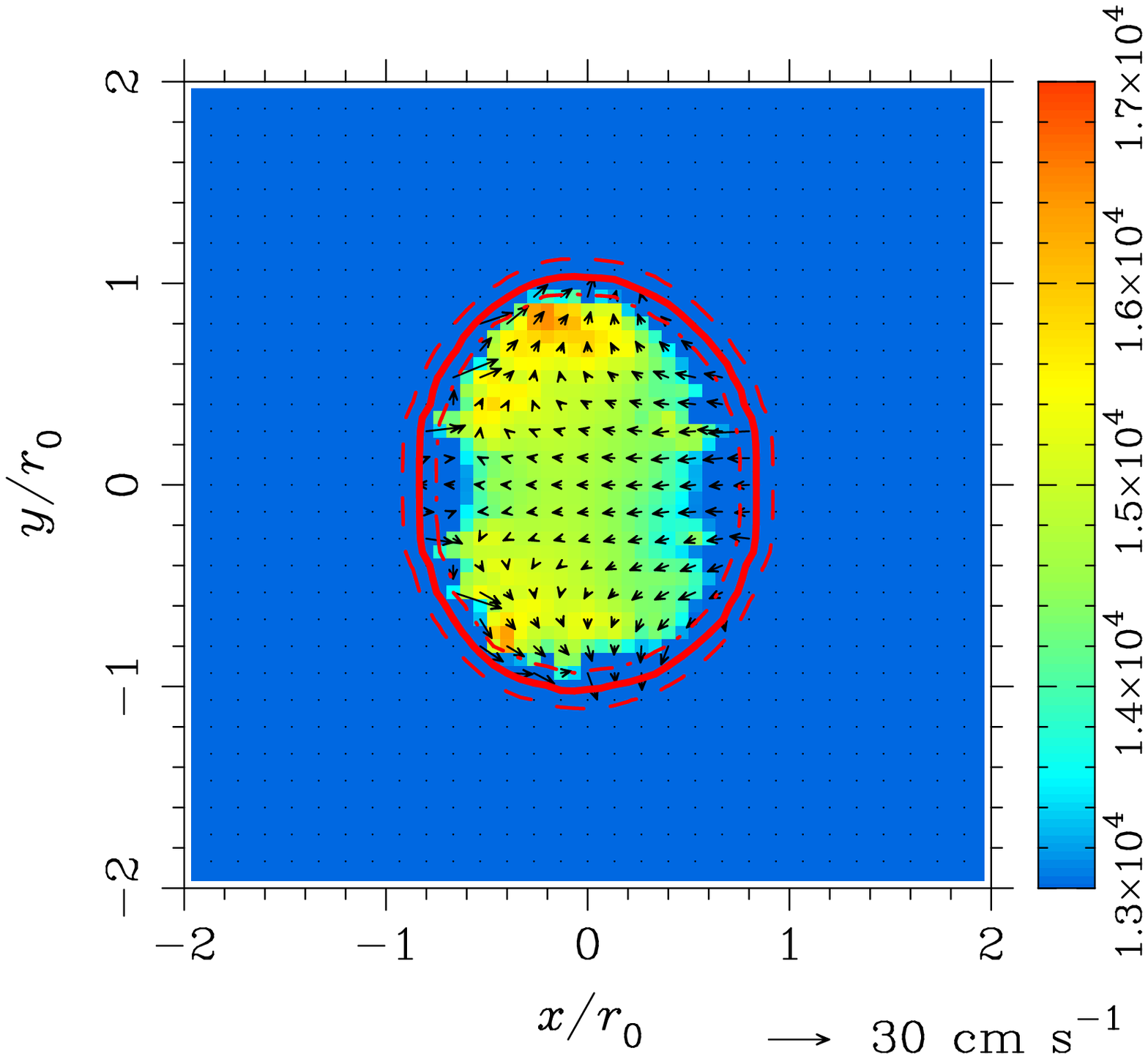}
	\end{minipage} &
	\hspace{6mm}
	\begin{minipage}{.35\linewidth}
		\begin{center}(f) 2.2 msec\end{center}
		\vspace{-5mm}
		\includegraphics[width=\linewidth, angle=-90.]{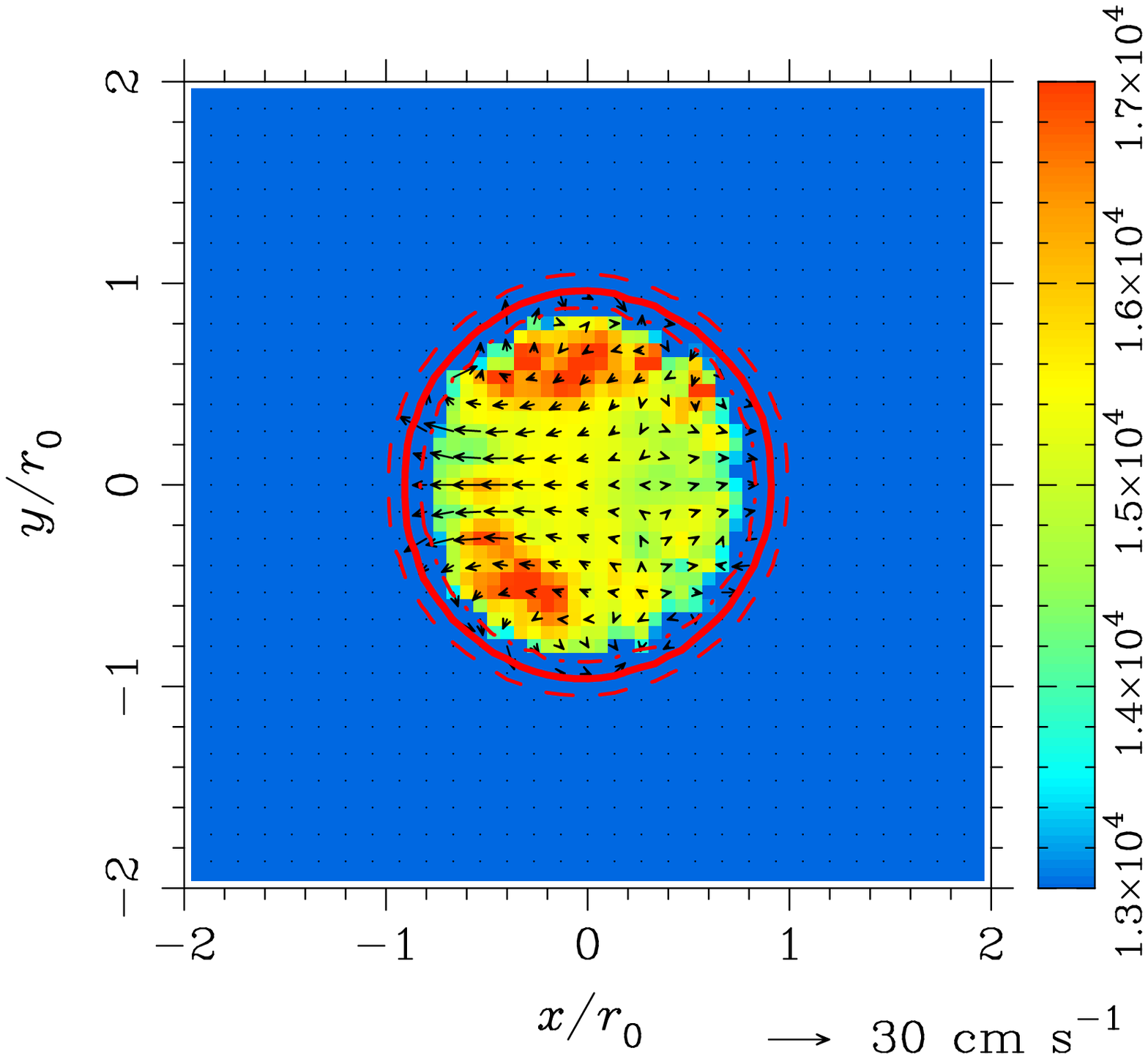}
	\end{minipage} \\
\end{tabular}
\caption{Time evolution of molten droplet exposed to the gas flow. The gas flow comes from the left side of panels. The initial droplet radius is $r_0 = 500 \, {\rm \mu m}$ (it corresponds to the Weber number $W_e = 0.5$).}
\label{fig:gasdrag_a00500_timeevo}
\end{figure}

Figure \ref{fig:gasdrag_a00500_timeevo} shows the time evolution of the molten droplet exposed to the gas flow. The initial droplet radius $r_0$ is set as $500 \, {\rm \mu m}$. Since this condition corresponds to the Weber number $W_e = 0.5$, the droplet does not undergo the fragmentation (see \S\ref{sec:initial_parameter}). A panel (a) shows the initial condition for the calculation. The gas flow is coming from the left side to the right side. The panel (b) shows the results after $0.3 \times 10^{-3} \, {\rm sec}$. The left side of the droplet is directly facing the gas flow, so the hydrostatic pressure at the left side becomes higher than that of the opposite side. The fluid elements at the surface layer are blown to the downstream, but the inner velocity turns to upstream of the gas flow because the apparent gravitational acceleration takes place in our coordinate system. The droplet continues to deform more and more, and after $1.0 \times 10^{-3} \, {\rm sec}$, the magnitude of the deformation becomes maximum. After that, the droplet begins to recover its shape to the sphere due to the surface tension. The panel (e) shows the shape in the middle of returning to a sphere, and the recovery motion is all but almost over at the panel (f). The droplet would repeat the deformation by the ram pressure of the gas flow and the recovery motion by the surface tension until the viscosity dissipates the internal motion of the droplet. 

\hspace{1cm}{\bf [Figure \ref{fig:gasdrag_a00500_timeevo}]}

\subsection{Deformation at Steady State}
As shown in Fig. \ref{fig:gasdrag_a00500_timeevo}, the droplet shows the vibrational motion: deformation by the ram pressure of the gas flow and recovery motion by the surface tension. It is expected that the vibrational motion ceases by the viscous dissipation and finally settles in a steady state. We plot the time evolution of the axial ratio C/B, where C and B are the minor axis and middle-length axis of the droplet, respectively. In the case of Fig. \ref{fig:gasdrag_a00500_timeevo}, it can be approximately considered that C and B are the droplet radii along the $x$- and $y$-axes, respectively (see \S\ref{sec:with_chondrule}). Therefore, C/B is an indicator of the droplet deformation, for example, ${\rm C/B} = 1$ approximately indicates a sphere and smaller C/B indicates a larger deformation. 

\begin{figure}[]
\center
\begin{tabular}{cc}
	\vspace{5mm}
	\begin{minipage}{.45\linewidth}
		\begin{center}(a) $r_0 = 100 \, {\rm \mu m}$ ($W_e = 0.1$)\end{center}
		\vspace{-3mm}
		\includegraphics[width=\linewidth, angle=-0.]{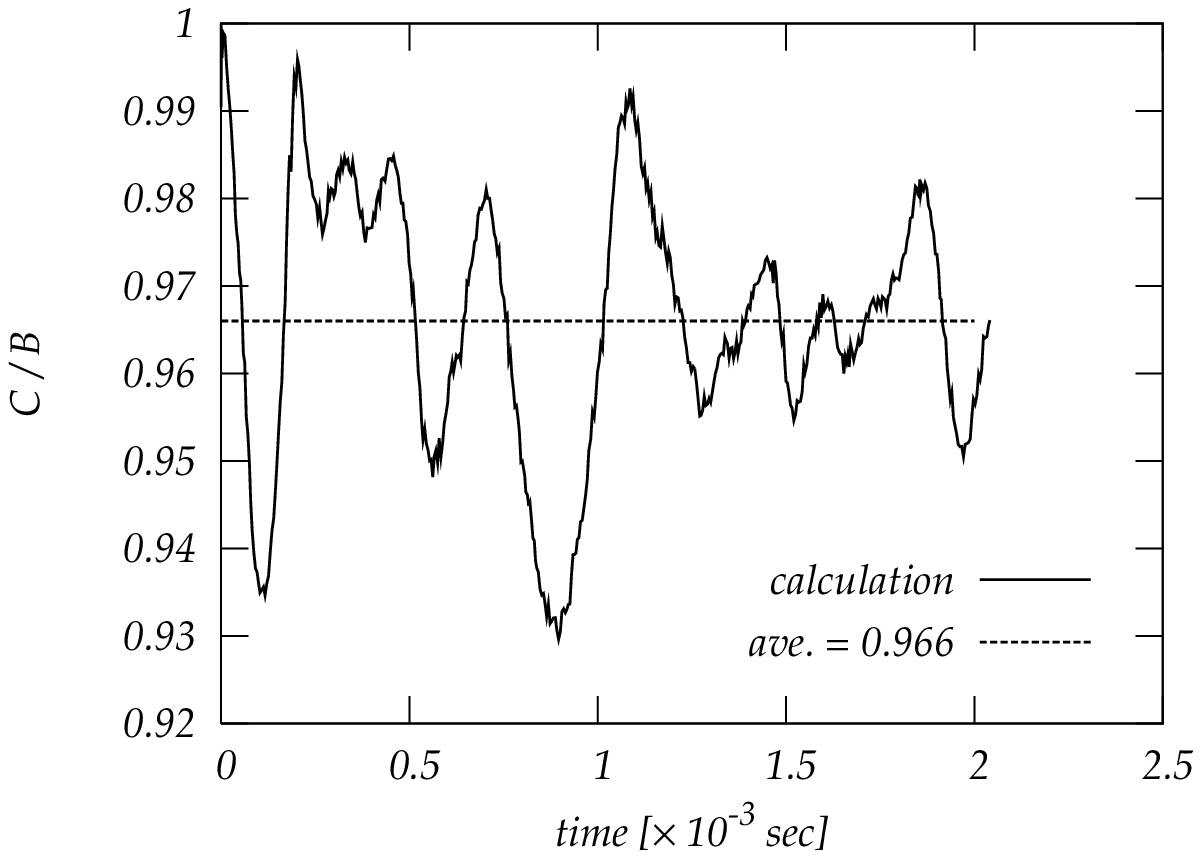}
	\end{minipage} &
	\begin{minipage}{.45\linewidth}
		\begin{center}(d) $r_0 = 1000 \, {\rm \mu m}$ ($W_e = 1.0$)\end{center}
		\vspace{-3mm}
		\includegraphics[width=\linewidth, angle=-0.]{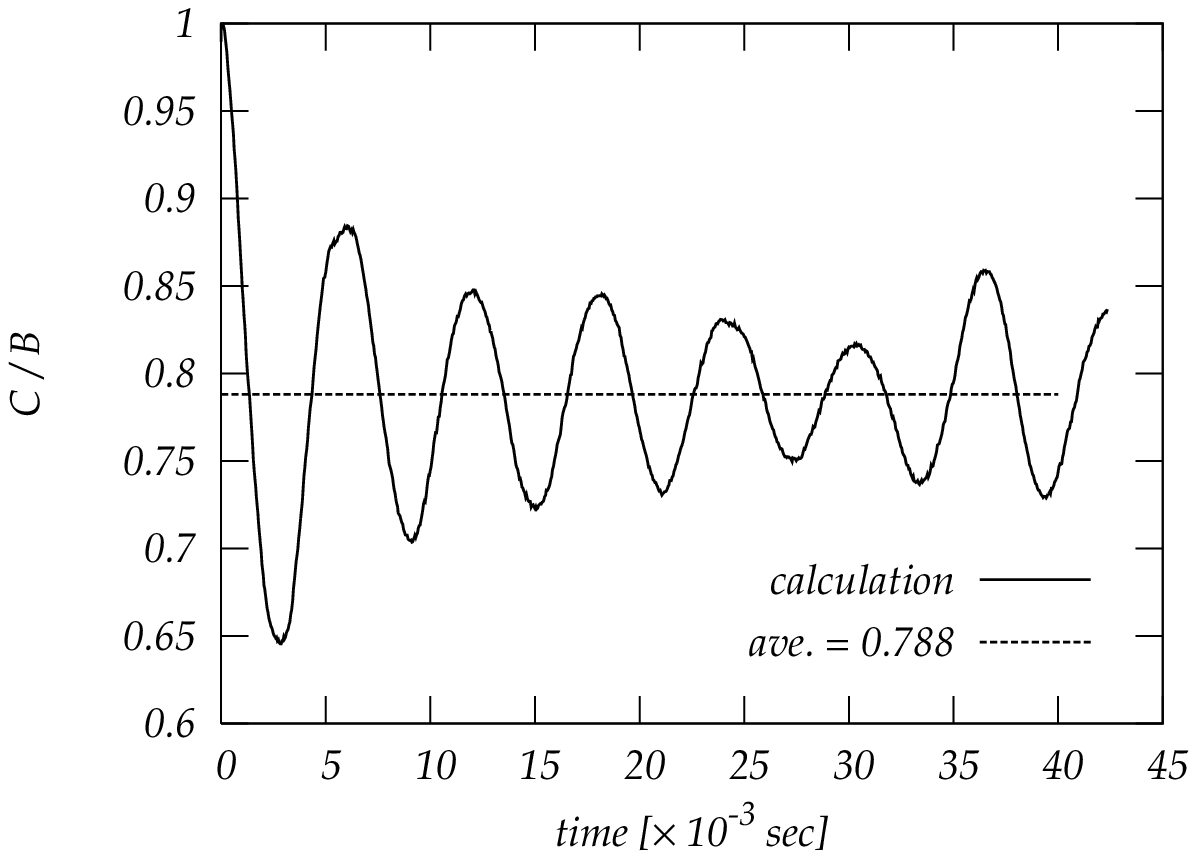}
	\end{minipage} \\
	\vspace{5mm}
	\begin{minipage}{.45\linewidth}
		\begin{center}(b) $r_0 = 200 \, {\rm \mu m}$ ($W_e = 0.2$)\end{center}
		\vspace{-3mm}
		\includegraphics[width=\linewidth, angle=-0.]{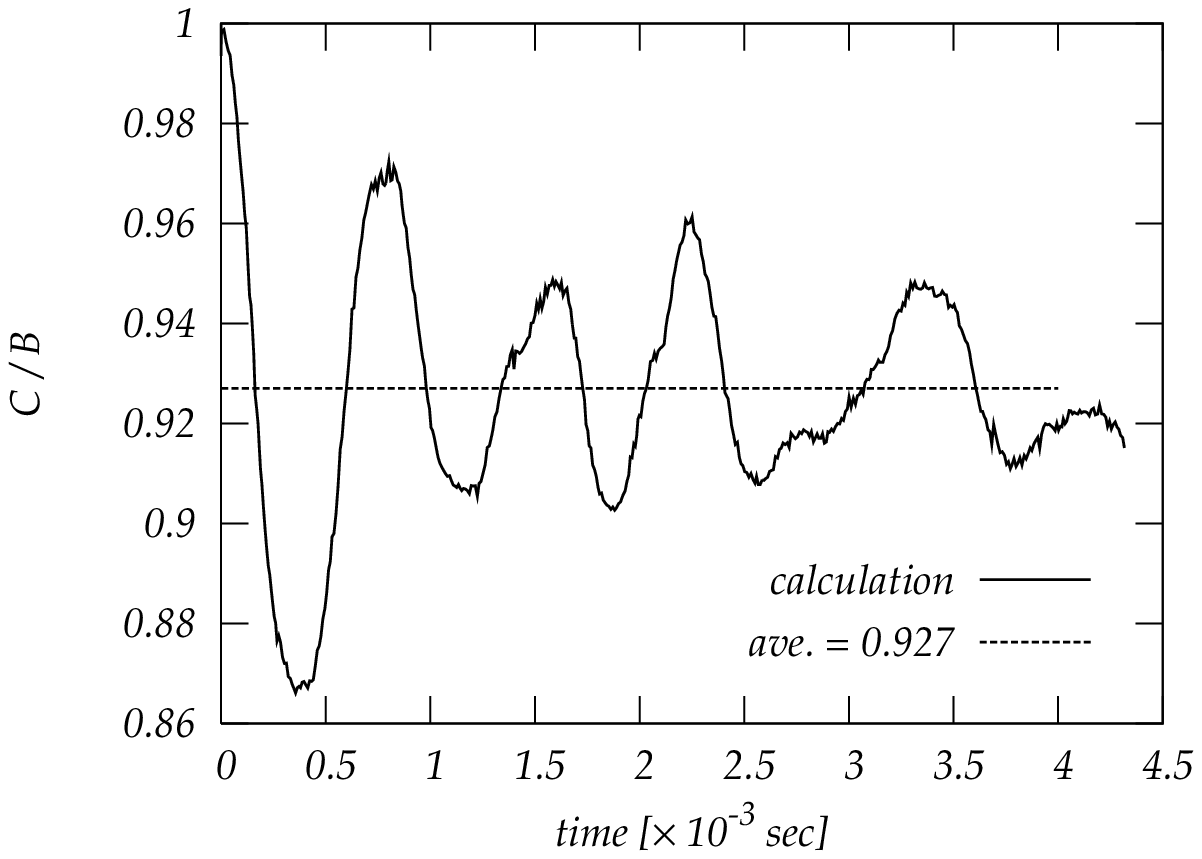}
	\end{minipage} &
	\begin{minipage}{.45\linewidth}
		\begin{center}(e) $r_0 = 2000 \, {\rm \mu m}$ ($W_e = 2.0$)\end{center}
		\vspace{-3mm}
		\includegraphics[width=\linewidth, angle=-0.]{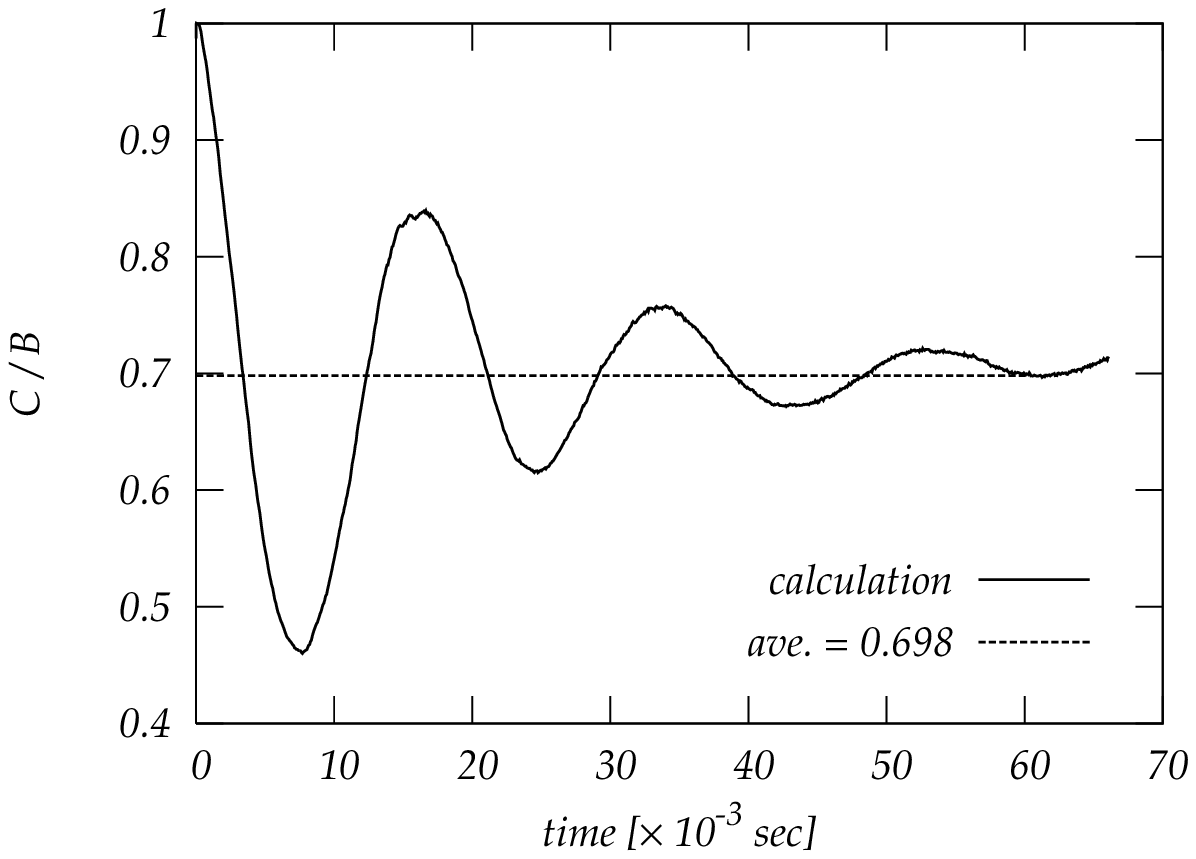}
	\end{minipage} \\
	\vspace{5mm}
	\begin{minipage}{.45\linewidth}
		\begin{center}(c) $r_0 = 500 \, {\rm \mu m}$ ($W_e = 0.5$)\end{center}
		\vspace{-3mm}
		\includegraphics[width=\linewidth, angle=-0.]{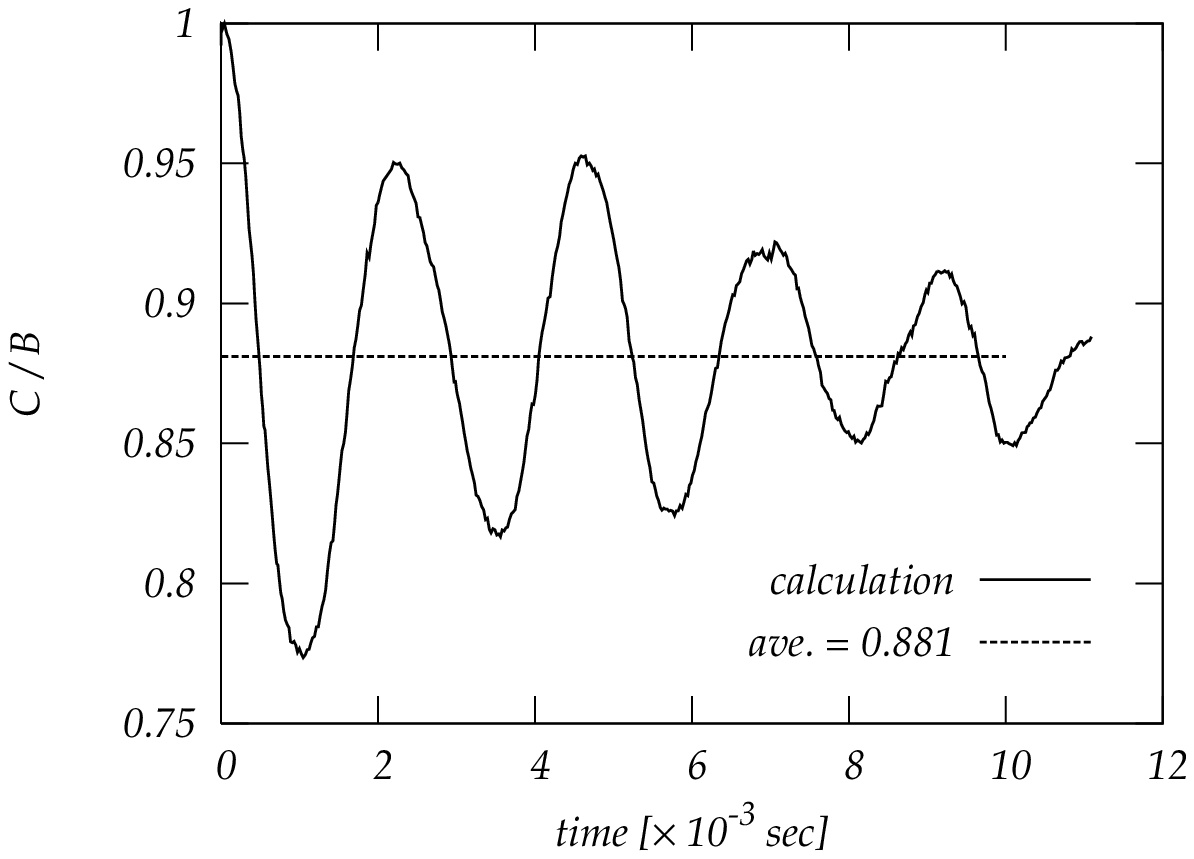}
	\end{minipage} &
	\begin{minipage}{.45\linewidth}
		\begin{center}(f) $r_0 = 5000 \, {\rm \mu m}$ ($W_e = 5.0$)\end{center}
		\vspace{-3mm}
		\includegraphics[width=\linewidth, angle=-0.]{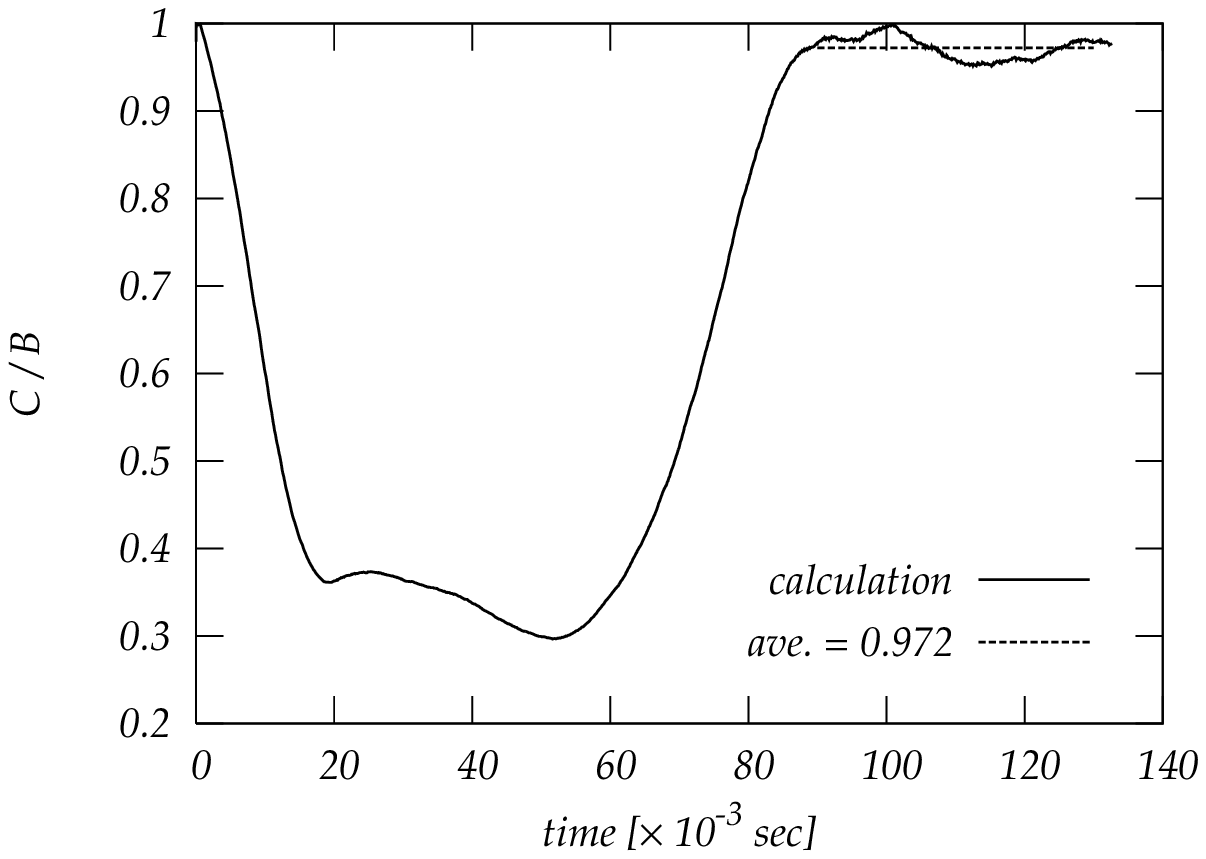}
	\end{minipage} \\
\end{tabular}
\caption{Vibrational motions of molten droplets for various radii: the deformation by the ram pressure and the recovery motion by the surface tension. The horizontal axis is the time since the ram pressure begins to affect the droplet and the vertical axis is the axial ratio of the droplet C/B (see text). The solid curves are the computational results and the dashed lines indicate the time-averaged values.}
\label{fig:time_CB}
\end{figure}

Figure \ref{fig:time_CB} shows the time evolutions of the axial ratios C/B for various radii of the droplets. The horizontal axis is the time after the ram pressure begins to act on the droplet. The solid curves are the computational results and the dashed lines indicate the time-averaged values defined as
\begin{equation}
\langle {\rm C/B} \rangle \equiv \frac{ \int ({\rm C/B}) dt }{ \int dt }, 
\label{eq:time_averaged_C/B}
\end{equation}
where the interval of the integration over $t$ is from the left edge to the right edge of each dashed line. Although some time evolutions in Fig. \ref{fig:time_CB} indicate that the vibrational motions do not terminate at the end of each calculation, the time-averaged values seem to well represent the axial ratio C/B at the final steady states. Then, we compare these time-averaged axial ratios with the linear solution derived by Sekiya et al. (2003), which describes the deformation of the droplet at the steady state. In order to compare the linear solution with our calculations in the same manner, we need to obtain the moment of inertia for the linear solution. In appendix \ref{appendix:moment_of_inertia}, we briefly summarize how to calculate it. 

\hspace{1cm}{\bf [Figure \ref{fig:time_CB}]}

\begin{figure}[]
\center
\includegraphics[scale=1., angle=0.]{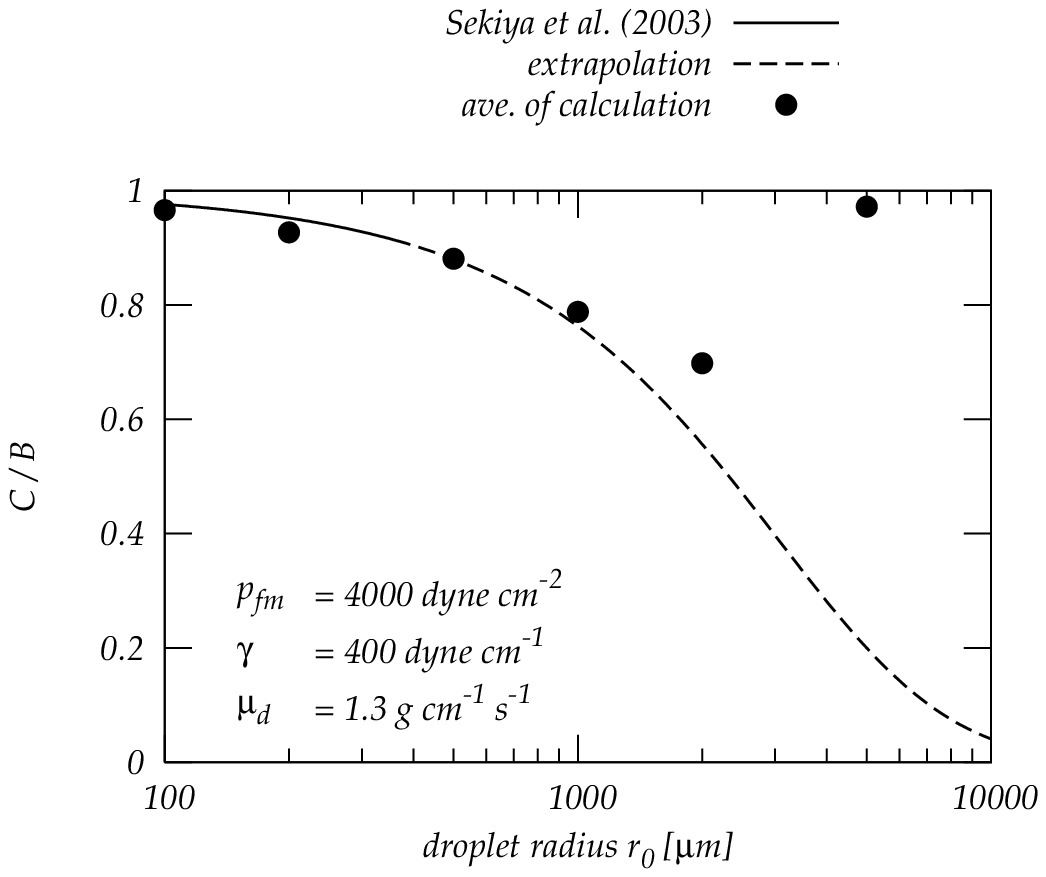}
\caption{Time-averaged values of axial ratios C/B as a function of the initial droplet radius $r_0$ (filled circles). The linear solutions derived by Sekiya et al. (2003) are also displayed (solid curve). The dashed curve indicates a simple extrapolation of the linear solutions.}
\label{fig:comp_sekiya}
\end{figure}

Figure \ref{fig:comp_sekiya} shows the comparisons of the time-averaged axial ratios (filled circles) with the linear solution (solid curve). The horizontal axis is the initial droplet radius $r_0$ and the vertical axis is the axial ratio C/B. Under the calculation conditions we adopted here, the linear solution cannot be applied for $r_0 \ga 358 \, {\rm \mu m}$ because the Reynolds number $R_e$ exceeds unity (Sekiya et al. 2003). The dashed curve is a simple extrapolation of the linear solution. It is found that the time-averaged values $\langle {\rm C/B} \rangle$ seem to show a good agreement not only at the radius range in which the linear solution can be applied but also out of the range ($r_0 \la 1000 \, {\rm \mu m}$). If the radius exceeds about $1000 \, {\rm \mu m}$, the non-linear terms in the hydrodynamical equations appear in the simulation results. The result of $r_0 = 2000 \, {\rm \mu m}$, in which the vibrational motion almost terminates and the droplet shape settles in the steady state, shows the slightly different value of C/B comparing with the extrapolation of the linear solution. The panel (f) of Fig. \ref{fig:time_CB} shows that the droplet of $r_0 = 5000 \, {\rm \mu m}$ once deforms significantly at $2 \times 10^{-2} \, {\rm sec} \la t \la 6 \times 10^{-2} \, {\rm sec}$, then the surface tension has the shape recover. However, the case of the panel (f) does not undergo the vibrational motion around the time-averaged value $\langle {\rm C/B} \rangle$, unlike other cases. After the first recovery motion, the droplet shape does not deform largely again. At the end of this calculation, the droplet settles around ${\rm C/B} \simeq 0.9$ with a slightly large scatter. 

\hspace{1cm}{\bf [Figure \ref{fig:comp_sekiya}]}

\subsection{Possibility of Cavitation\label{sec:cavitation}}
In Fig. \ref{fig:comp_sekiya}, we found that the axial ratio C/B of the droplet of $r_0 = 5000 \, {\rm \mu m}$ departs from the extrapolation of the linear solution derived by Sekiya et al. (2003), in which the non-linear terms of the hydrodynamical equations as well as the surface deformation were neglected. Here, we show the effects of the non-linear terms in hydrodynamics of the droplet. 

\begin{figure}[]
\center
\begin{tabular}{cc}
	\vspace{3mm}
	\begin{minipage}{.35\linewidth}
		\begin{center}(a) 11 msec\end{center}
		\vspace{-5mm}
		\includegraphics[width=\linewidth, angle=-90.]{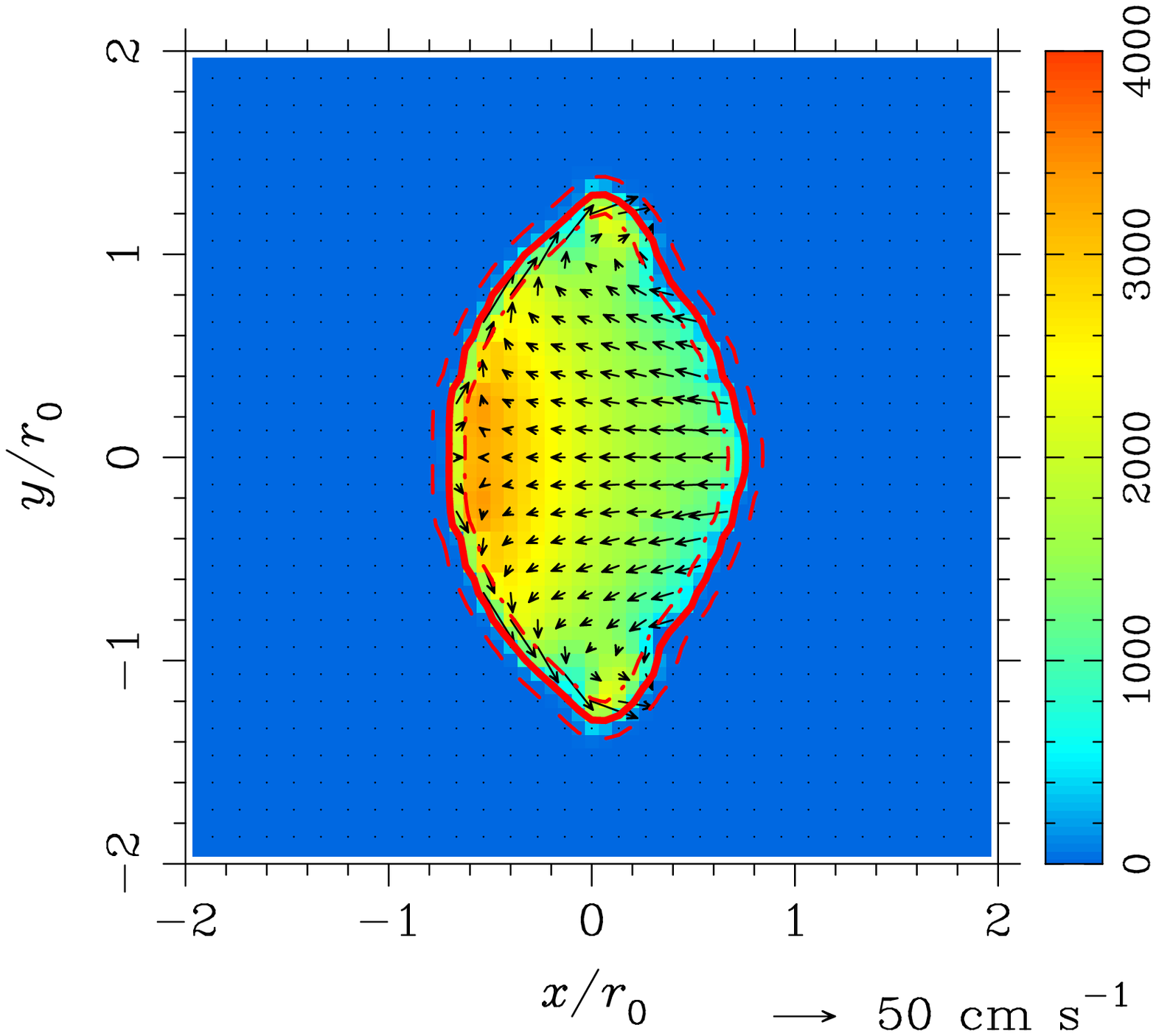}
	\end{minipage} &
	\hspace{6mm}
	\begin{minipage}{.35\linewidth}
		\begin{center}(d) 71 msec\end{center}
		\vspace{-5mm}
		\includegraphics[width=\linewidth, angle=-90.]{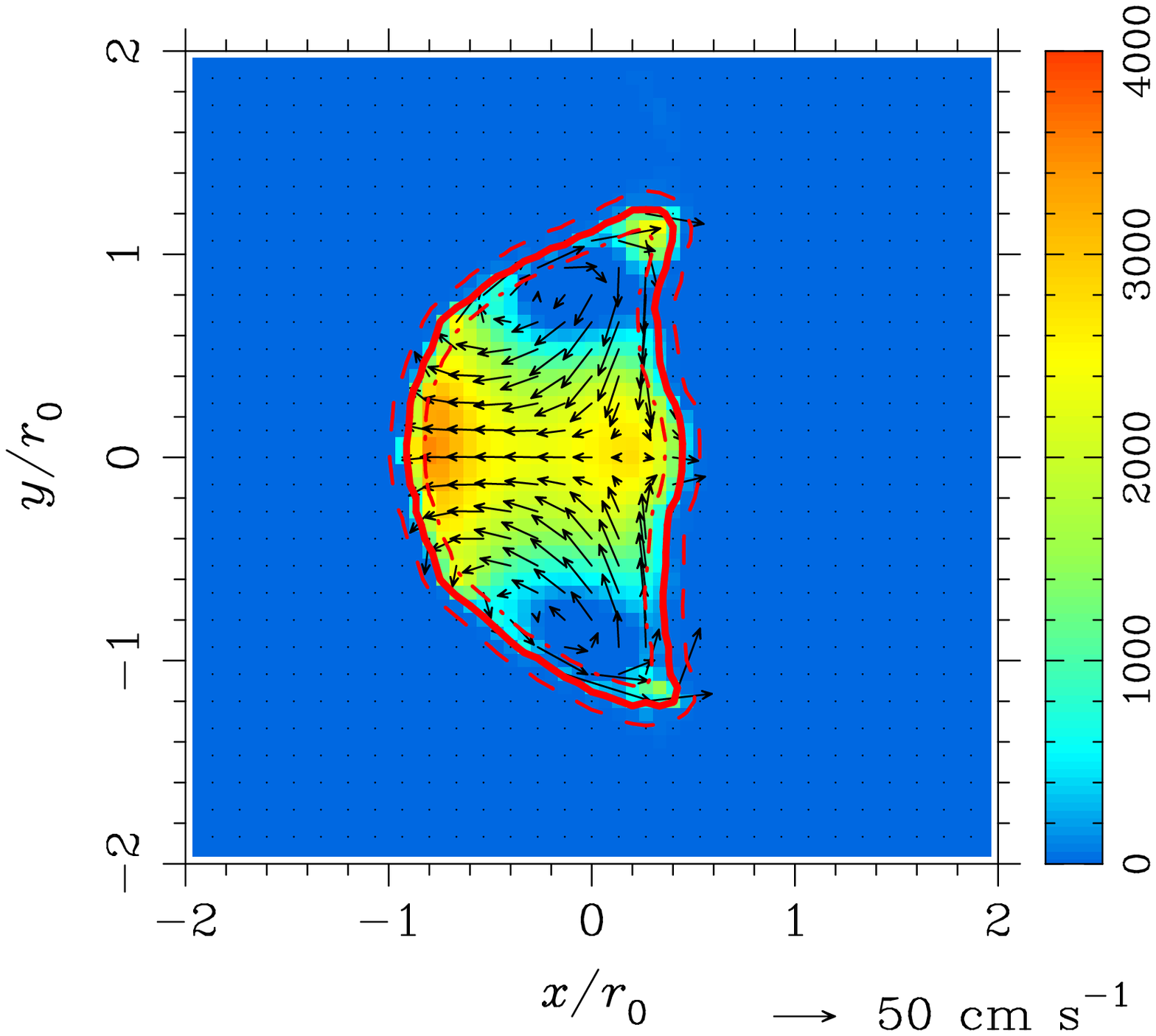}
	\end{minipage} \\
	\vspace{3mm}
	\begin{minipage}{.35\linewidth}
		\begin{center}(b) 28 msec\end{center}
		\vspace{-5mm}
		\includegraphics[width=\linewidth, angle=-90.]{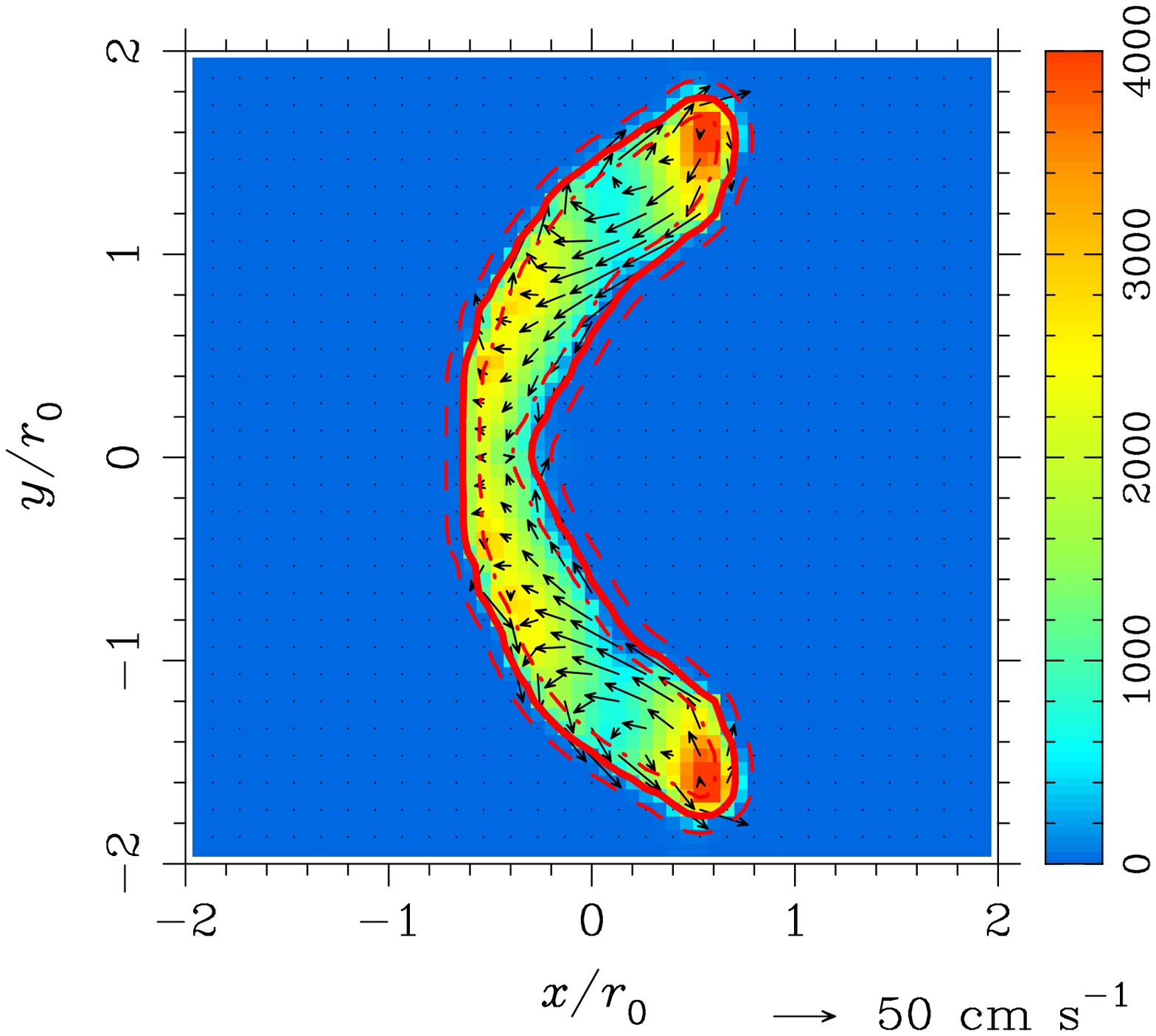}
	\end{minipage} &
	\hspace{6mm}
	\begin{minipage}{.35\linewidth}
		\begin{center}(e) 79 msec\end{center}
		\vspace{-5mm}
		\includegraphics[width=\linewidth, angle=-90.]{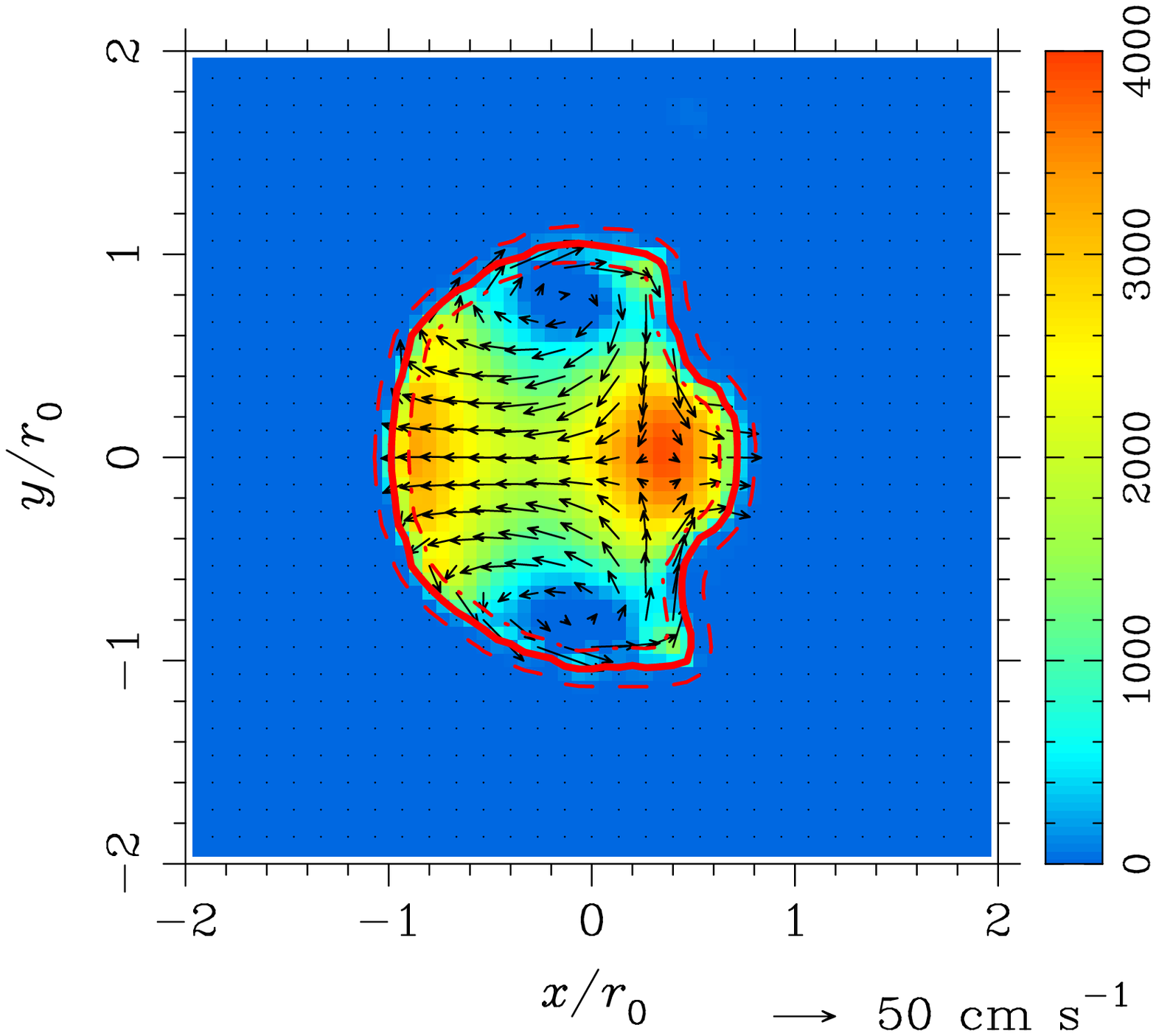}
	\end{minipage} \\
	\vspace{3mm}
	\begin{minipage}{.35\linewidth}
		\begin{center}(c) 62 msec\end{center}
		\vspace{-5mm}
		\includegraphics[width=\linewidth, angle=-90.]{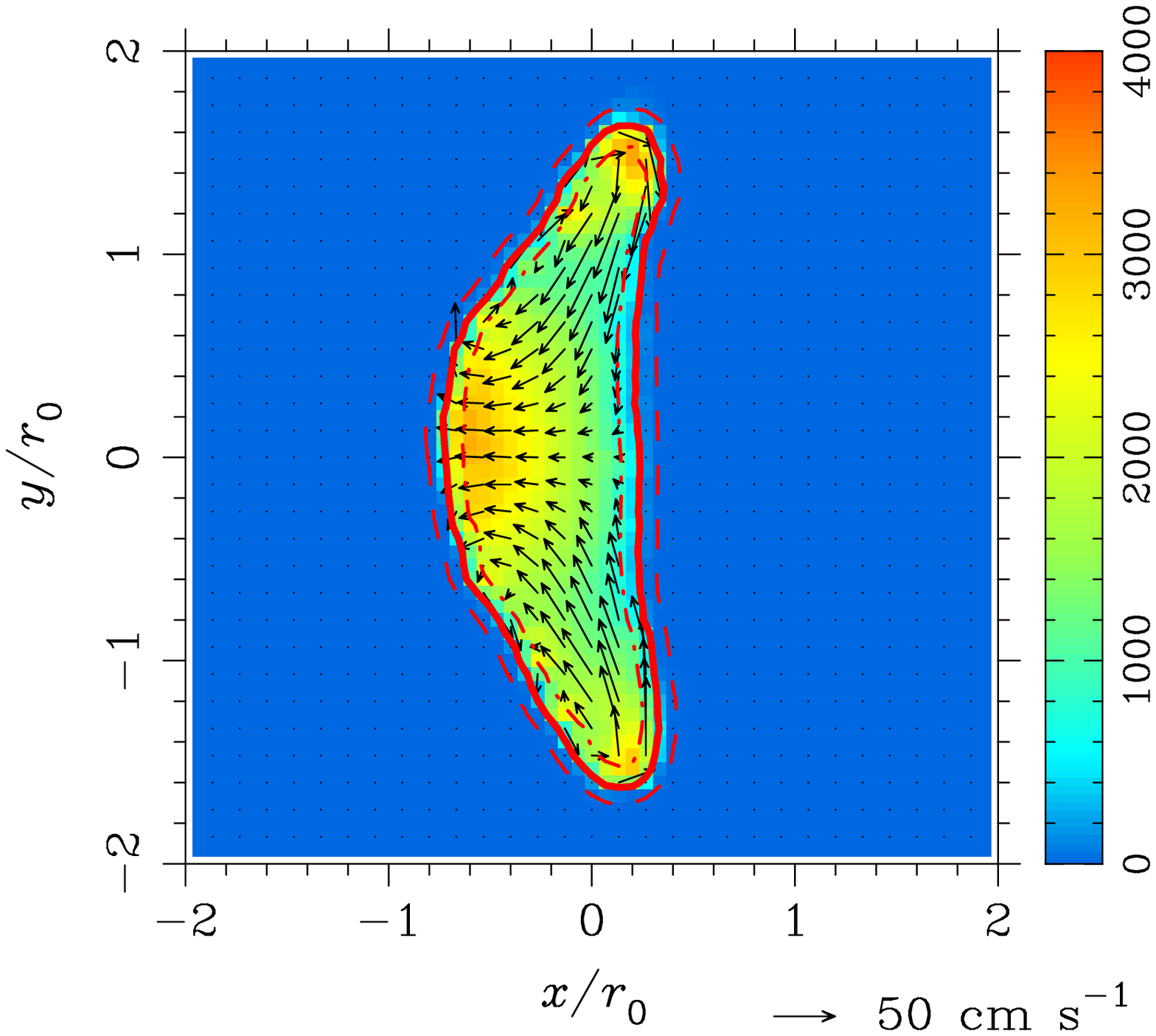}
	\end{minipage} &
	\hspace{6mm}
	\begin{minipage}{.35\linewidth}
		\begin{center}(f) 107 msec\end{center}
		\vspace{-5mm}
		\includegraphics[width=\linewidth, angle=-90.]{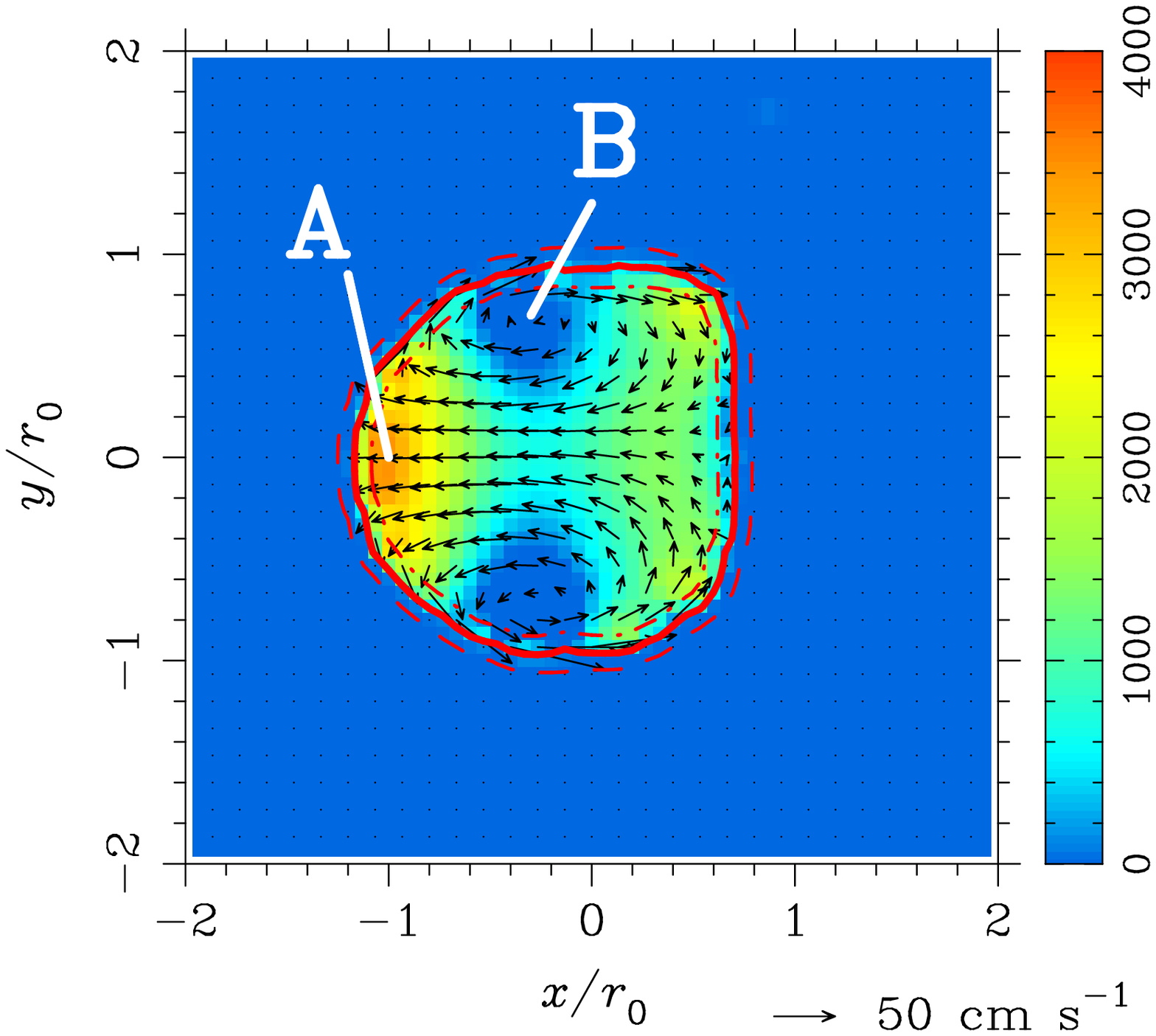}
	\end{minipage} \\
\end{tabular}
\caption{Same as Fig. \ref{fig:gasdrag_a00500_timeevo} except for $r_0 = 5000 \, {\rm \mu m}$ ($W_e = 5$).}
\label{fig:gasdrag_a05000_timeevo}
\end{figure}

Figure \ref{fig:gasdrag_a05000_timeevo} shows the time evolution of the droplet for $r_0 = 5000 \, {\rm \mu m}$. It is found that the droplet deforms significantly around $t = 2.8 \times 10^{-2} \, {\rm sec}$ (panel b). The large amount of the fluid spreads out to the perpendicular direction of the gas flow ($y$- and $z$-directions). Although the center of the flatten droplet becomes very thin, it does not fragment directly. The fluid blown out at once returns to the back of the droplet and gathers at the center (panels c and d). The center of the droplet at which the fluid elements gather becomes high pressure and the pressure gradient force pushes the fluid elements toward right in the panel. As a result, the back of the droplet pops out (panel e). The bump gradually disappears with time, and finally, the droplet settles on almost the steady state with the internal flow largely circulating all around the droplet (panel f). In the phase, the hydrostatic pressure is high at the part which is directly facing the gas flow and extremely low around the center of eddies of the circulating internal fluid motion. The final pressure distribution is qualitatively different from the linear solution derived by Sekiya et al. (2003), in which the pressure at the front of the droplet (pointed by ``A" in the panel) is lower than that at the side (pointed by ``B"). 

Here, we would like to point out the behavior of the low-pressure region at the center of the eddies (pointed by ``B" in the panel f). Generally, it is considered that the boiling (or vaporization) would take place in any liquids where the vapor pressure of the liquid exceeds its hydrostatic pressure. The vapor pressure of forsterite, which is one of the main components of chondrules, is about $1 \, {\rm dyne \, cm^{-2}}$ at $1850 \, {\rm K}$ (Miura et al. 2002). On the contrary, since the hydrostatic pressure at the center of eddies is almost zero, the boiling might occur at the center of eddies. We call this phenomenon the cavitation. It indicates that the droplet might be cut off at the place where bubbles are generated by the cavitation. Therefore, it is thought that the droplet fragments into smaller pieces in the real situation. 

Miura et al. (2002) discussed that the ram pressure of the gas flow produces the high-pressure environment in the molten droplet, and the pressure is high enough to suppress the boiling in it when the gas frictional heating is strong enough to melt the silicate dust particles. However, they did not take into account the hydrodynamics of the molten droplet. The fragmentation by cavitation might be a new mechanism to explain the maximum size of chondrules (see \S\ref{sec:chondrule_maximum_size}) like as the other mechanisms; the fragmentation due to the gas flow directly (Susa \& Nakamoto 2002) or the stripping of the liquid surface in the gas flow (Kato et al. 2006, Kadono \& Arakawa 2005). To simulate the behavior of the bubbles generated in the liquids should be the challenging task, but we think that such kind of simulations are very important. It might be related with the existence of holes in some chondrules (e.g., Kondo et al. 1997). It should be taken into consideration in the future work. 

\hspace{1cm}{\bf [Figure \ref{fig:gasdrag_a05000_timeevo}]}

\subsection{Fragmentation}
\begin{figure}[]
  \center
  \begin{tabular}{cc}
    \begin{minipage}{.4\linewidth}
      \begin{center}(a) 0.068 sec\end{center}
      \includegraphics[width=\linewidth, angle=-0.]{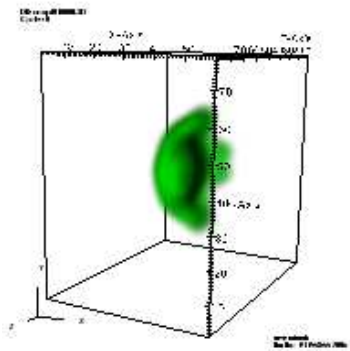}
    \end{minipage} &
    \hspace{1mm}
    \begin{minipage}{.4\linewidth}
      \begin{center}(b) 0.143 sec\end{center}
      \includegraphics[width=\linewidth, angle=-0.]{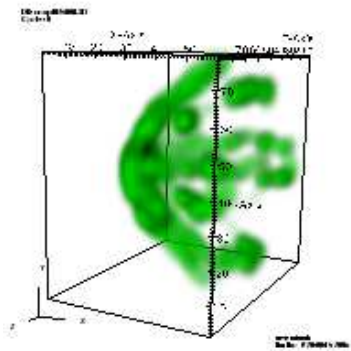}
    \end{minipage} \\
  \end{tabular}
\caption{Three-dimensional views of the fragmentation of molten droplet. The initial droplet radius is $r_0 = 2 \, {\rm cm}$ ($W_e = 20$).}
\label{fig:gasdrag_a200003d_timeevo}
\end{figure}

For the droplet with larger size in which the surface tension cannot keep the droplet shape against the gas ram pressure, the fragmentation will occur. Figure \ref{fig:gasdrag_a200003d_timeevo} shows the three-dimensional views of the break-up droplet. The droplet radius is $r_0 = 2 \, {\rm cm}$, which corresponds to $W_e = 20$. In this case, we take a wider computational domain ($-3 < x / r_0 < 3$ and $-4 < y / r_0, z / r_0 < 4$) than previous simulations in order to treat the break-up phenomenon. The computational grid points are $60 \times 80 \times 80$, so the spatial resolution is worse than that of the previous cases. The gas flow comes from the left side of the view along the $x$-axis. The panel (a) shows the droplet shape just before the fragmentation. It is found that the droplet surface which is directly facing to the gas flow is stripped off backward. The panel (b) is just after the fragmentation. We found that the parent droplet breaks up to many smaller pieces. If these pieces collide again behind the parent droplet, the compound chondrules, which are thought to have collided two independent chondrules, might be formed. 

In Fig. \ref{fig:gasdrag_a200003d_timeevo}, the fragmentation does not seem to be axis-symmetric. The reason is thought to be the directional splitting technique for multi-dimensions in the Cartesian coordinate. Therefore, we would not be able to quantitatively discuss the sizes and number of the fragments. In order to do that in the Cartesian coordinate system, we might need higher spatial resolutions. 

\hspace{1cm}{\bf [Figure \ref{fig:gasdrag_a200003d_timeevo}]}

\section{Discussions}

\subsection{Maximum Size of Chondrules\label{sec:chondrule_maximum_size}}
Susa \& Nakamoto (2002) suggested that the fragmentation of the droplets in high-velocity gas flow limits the sizes of chondrules (upper limit). They considered the balance between the surface tension and the inhomogeneity of the ram pressure acting on the droplet surface, and derived the maximum size of molten silicate dust particles above which the droplet would be destroyed by the ram pressure of the gas flow using an order of magnitude estimation. In their estimation, they adopted the experimental data in which the droplets suddenly exposed to the gas flow fragment for $W_e \ga 6$ (Bronshten 1983, p.96). This results into the fragmentation of droplet for $r_0 \ga 6000 \, {\rm \mu m}$ if we adopt our calculation conditions ($p_{\rm fm} = 4000 \, {\rm dyne \, cm^{-2}}$ and $\gamma = 400 \, {\rm dyne \, cm^{-1}}$). 

On the contrary, our simulations showed the possibility that the droplets fragment through the cavitation even if the radius is smaller than the maximum size derived by Susa \& Nakamoto (2002) (see \S\ref{sec:cavitation}). The cavitation would occur when the pressure gradient force cannot support the fluid motion to circulate around the eddy. Here, we consider the condition that the cavitation would take place using a simple order of magnitude estimation. The pressure gradient force per unit volume can be estimated as $f_{\rm pres} \sim p / ( r_0 / 2 )$, where $r_0 / 2$ indicates the radius of the eddy. On the other hand, the centrifugal force for the fluid element circulating around the eddy can be given as $f_{\rm cent} \sim \rho_{\rm d} v_{\rm circ}^2 / ( r_0 / 2 )$, where $v_{\rm circ}$ is the rotational velocity. Balancing the pressure gradient and the centrifugal force around the eddy, we have
\begin{equation}
\frac{p}{r_0 / 2} \sim \rho_{\rm d} \frac{v_{\rm circ}^2}{r_0 / 2}.
\label{eq:centrifugal_pressure_balance}
\end{equation}
Substituting $p = 2 \gamma / r_0$ and $v_{\rm circ} \simeq v_{\rm max} = 0.112 p_{\rm fm} r_0 / \mu_{\rm d}$ (Sekiya et al. 2003), we obtain
\begin{equation}
r_0 \sim \bigg( \frac{ 2 \gamma \mu_{\rm d}^2}{ 0.112^2 \rho_{\rm d} p_{\rm fm}^2} \bigg)^{1/3}. 
\label{eq:centrifugal_pressure_balance2}
\end{equation}
This equation gives the critical radius of the droplet above which the cavitation might occur in the center of the eddy and result into the fragmentation. 

\begin{figure}[]
\center
\includegraphics[scale=.7, angle=0.]{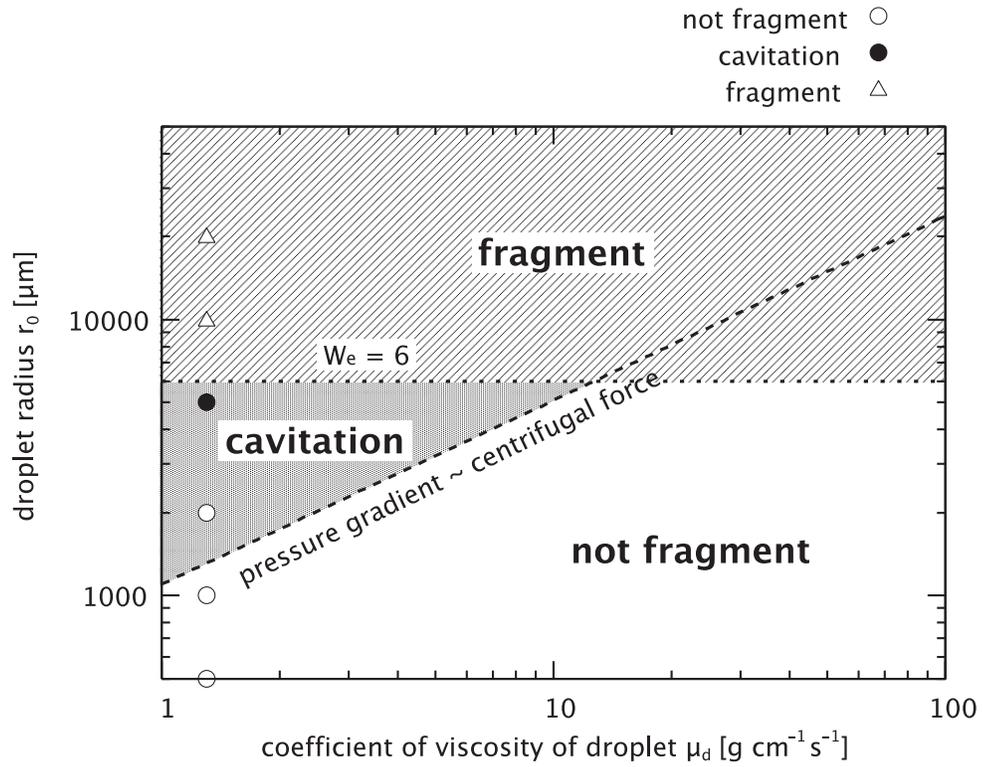}
\caption{Fates of the dust particles when they melt in the high-velocity gas flow. The horizontal axis is the coefficient of viscosity of droplet $\mu_{\rm d}$ and the vertical axis is the droplet radius $r_0$. We assume $p_{\rm fm} = 4000 \, {\rm dyne \, cm^{-2}}$, $\gamma = 400 \, {\rm dyne \, cm^{-1}}$, and $\rho_{\rm d} = 3 \, {\rm g \, cm^{-3}}$, respectively. Our simulation results are also shown (symbols).}
\label{fig:pressure_centri}
\end{figure}

Figure \ref{fig:pressure_centri} shows the two critical radii as a function of the viscosity; the radius for $W_e = 6$ (dotted line) and the radius at which the pressure gradient force balances with the centrifugal force (dashed line). We assume $p_{\rm fm} = 4000 \, {\rm dyne \, cm^{-2}}$, $\gamma = 400 \, {\rm dyne \, cm^{-1}}$, and $\rho_{\rm d} = 3 \, {\rm g \, cm^{-3}}$, respectively. Our simulation results are also shown; not fragment (circles), cavitation would take place (filled circles), and fragmentation (triangle), respectively. It is found that there is a critical value for the viscous coefficient of the molten droplet $\mu_{\rm cr}$ below which the cavitation would take place at a certain region in radius, even if $W_e$ is not so large that the fragmentation is not expected to occur. The critical viscous coefficient $\mu_{\rm cr}$ is calculated as
\begin{eqnarray}
\mu_{\rm cr} &\sim& \bigg( \frac{ 6^3 \cdot 0.112^2 \gamma^2 \rho_{\rm d} }{ 2 p_{\rm fm} } \bigg)^{1/2} \nonumber \\
&=& 12.8 \, \bigg( \frac{ p_{\rm fm} }{ 4000 \, {\rm dyne \, cm^{-2}} } \bigg)^{-1/2}
\bigg( \frac{ \gamma }{ 400 \, {\rm dyne \, cm^{-1}} } \bigg) \, {\rm g \, cm^{-1} \, s^{-1}}.
\label{eq:critical_viscosity}
\end{eqnarray}

Susa \& Nakamoto (2002) clearly showed the maximum sizes of molten droplets in the post-shock region, $a_{\rm cr}$, as a function of the initial gas density $\rho_0$ and the shock velocity $v_{\rm s}$ (see Fig. 1 in their paper). They also compared $a_{\rm cr}$ with the shock conditions for the chondrule formation (in which the chondrule precursor dust particles can melt and do not evaporate completely). In their figure, they assumed that the droplet will be destroyed by the gas ram pressure when $W_e > 6$. However, if the droplet is so heated that the droplet viscosity exceeds the critical value $\mu_{\rm cr}$, even the smaller droplet would be destroyed by the cavitation. From above speculations, it is expected that BO chondrules (which are believed to have melted totally and experienced high temperature, e.g., Hewins \& Radomsky 1990) are smaller than POP chondrules (which are thought to have melted partially). On the contrary, there are no obvious size differences between BO and POP chondrules (e.g., Rubin 1989). This discrepancy might be explained by that the precursor dust particles were not so large. If the maximum precursor size is smaller than about $1000 \, {\rm \mu m}$, the cavitation would not take place for $\mu \ga 1 \, {\rm g \, cm^{-1} \, s^{-1}}$ in Fig. \ref{fig:pressure_centri}. Another possibility is that the viscosity of the molten droplets have not been below the critical value $\mu_{\rm cr}$, above which the cavitation would not take place. Namely, if the dust parameters do not enter the region of ``cavitation" in Fig. \ref{fig:pressure_centri}, the size-dependence in chondrule sizes would not appear by the cavitation. In order to investigate above issue, we might have to consider not only the hydrodynamical effect but also the thermal evolutions of the molten droplets in order to discuss the maximum sizes. The thermal evolution of the droplets is beyond the scope of this paper. We would like to investigate this issue in the future work.


\hspace{1cm}{\bf [Figure \ref{fig:pressure_centri}]}

\subsection{Final Shape of Droplet}
In this paper, we numerically simulated the hydrodynamics of molten droplets exposed to the gas flow. We assumed that the coefficient of viscosity is constant in the droplets, and the droplets melt completely and the physical conditions inside the droplets are homogeneous. On the contrary, the shapes of chondrules are considered to be determined around the re-solidification phase. Before the molten droplets re-solidify, the droplet temperature falls down and the coefficient of viscosity would change significantly. Moreover, if the droplets melt partially, the physical conditions inside the droplets should not be homogeneous. Therefore, we must discuss how we can consider above effects on the final shapes of chondrules. 

It is naturally considered that the viscosity of the droplet becomes larger as the droplet temperature falls down. In such a lower Reynolds number environment, the linear solution by Sekiya et al. (2003) is a better approximation to describe the external shape and internal flow of the droplet. According to their linear solution, the external shape of the droplet does not depend on the viscosity. It indicates that the external shape is determined mainly by the balance between the surface tension and the gas ram pressure. Therefore, it is considered that the droplet shape does not change significantly at the cooling phase in which the viscosity of the droplet becomes higher and higher. 

Another problem that we must take care for considering the droplet shapes is inhomogeneity of the physical conditions in the droplet. For example, if the dust particle is not heated enough, it would not melt completely and the solid lumps floating in the droplet would exist. In this case, the assumption of the completely molten droplet we adopted loses the validity. However, if the external shape is significantly affected by the solid lumps, the surface of chondrules should be irregular and cannot be approximated by smoothed shapes (e.g., spheres or ellipsoids). If considering conversely, it is thought that we can remove the above effect by considering only the chondrules which have smooth surfaces. In fact, there are some fraction of chondrules with smooth surfaces (Tsuchiyama et al. 2003). Such chondrules would have melted almost completely, at least the unmelted solid lumps do not affect the external shapes of droplets. 

However, the majority of chondrules has the porphyritic texture, which indicates that most of chondrule precursors did not completely melt in the heating event (Hewins \& Radomsky 1990). Moreover, iron sulfide inclusions are observed in natural chondrules with various forms (Uesugi et al. 2005). The unmelted clumps and the iron sulfide inclusions might disturb the internal flow of the molten silicate particles\footnote{Uesugi et al. (2005) considered that the trajectories of the iron sulfide inclusions in the silicate melts, however, they did not take into account the influence of the iron sulfide inclusions on the flow pattern of the ambient silicate melt. }. This problem is very complex and beyond the scope of this paper. We believe that this problem can be investigated in the future.

\subsection{Comparison with Chondrules\label{sec:with_chondrule}}
Tsuchiyama et al. (2003) studied three-dimensional shapes of chondrules using X-ray microtomography. They measured twenty chondrules with perfect shapes and smooth surfaces, which were selected from 47 chondrules separated from the Allende meteorite (CV3). The external shapes were approximated as three-axial ellipsoids with a-, b-, and c-axes (axial radii are A, B, and C (${\rm A} \ge {\rm B} \ge {\rm C}$), respectively) using the moments of inertia of the chondrules, where the rotation axes with the minimum and maximum moments correspond to the a- and c-axes, respectively. They found that (1) the shapes are diverse from oblate (${\rm A} \sim {\rm B} > {\rm C}$), general three-axial ellipsoid (${\rm A} > {\rm B} > {\rm C}$) to prolate chondrules (${\rm A} > {\rm B} \sim {\rm C}$), and (2) two groups can be recognized: oblate to prolate chondrules with $0.9 \la {\rm B/A} \la 1.0$ (group-A) and prolate chondrules with $0.74 \la {\rm B/A} \la 0.80$ (group-B). 

\begin{figure}[]
\center
\includegraphics[scale=1., angle=0.]{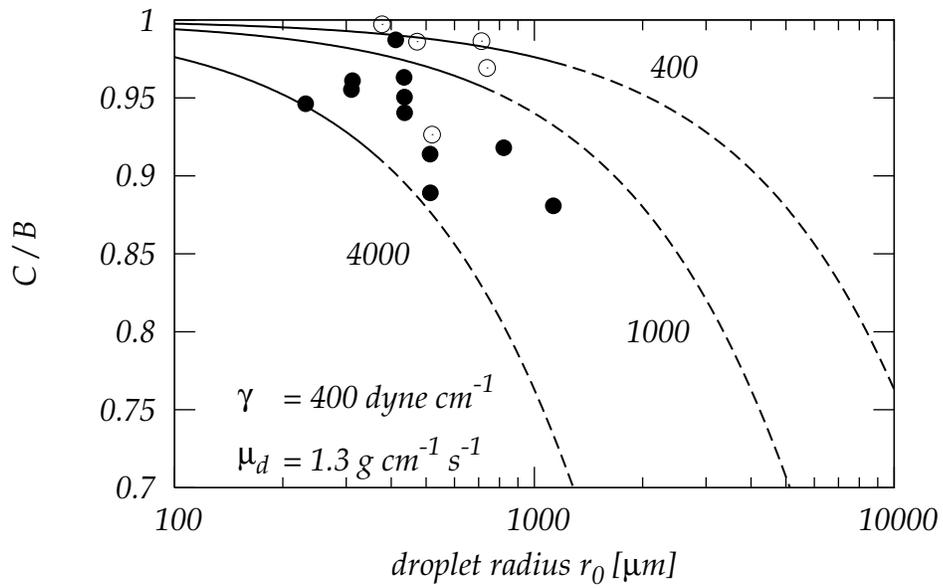}
\caption{The axial ratios C/B of group-A chondrules are plotted as a function of the chondrule radius; filled circles are oblate (${\rm B/A} > {\rm C/B}$) and open circles are others. The solid curves are the linear solutions for various ram pressures (Sekiya et al. 2003). The numbers in the panel indicate the values of $p_{\rm fm}$ in the unit of ${\rm dyne \, cm^{-2}}$. The dashed curves are simple extrapolations of the linear solutions.}
\label{fig:various_pfm}
\end{figure}

The axial ratio C/B of the group-A chondrules is about $0.9 \la {\rm C/B} \la 1.0$ and the radii are about from $200 \, {\rm \mu m}$ to $1000 \, {\rm \mu m}$ (A. Tsuchiyama, private communication). On the contrary, for the molten droplet exposed to the gas flow, C/B also depends on the momentum flux of the rarefied gas flow $p_{\rm fm}$. Figure \ref{fig:various_pfm} shows the axial ratio C/B for various values of $p_{\rm fm}$ ($ = 400$, $1000$, and $4000 \, {\rm dyne \, cm^{-2}}$) calculated by the linear solution derived by Sekiya et al. (2003). We also display the data of the group-A chondrules; filled circles are oblate (${\rm B/A} > {\rm C/B}$) and open circles are others. The radii of chondrules are estimated by $({\rm ABC})^{1/3}$. It is found that the data of group-A chondrules seem to distribute between the linear solutions for $p_{\rm fm} = 400 \, {\rm dyne \, cm^{-2}}$ and $p_{\rm fm} = 4000 \, {\rm dyne \, cm^{-2}}$. In other words, assuming that the axial ratio C/B of those chondrules were caused by the ram pressure of the gas flow when they melted, the expected ram pressure of the gas flow is about $400 - 4000 \, {\rm dyne \, cm^{-2}}$. This value is the typical value of the shock-wave heating model for chondrule formation. Therefore, it is strongly suggested that the group-A chondrules might have been formed in the high-velocity rarefied gas flow of the shock-wave heating. 

\hspace{1cm}{\bf [Figure \ref{fig:various_pfm}]}

However, it should be noted that some of those group-A chondrules have the axial ratios B/A which are not unity. This indicates that the shapes are not pure oblate but general three-axial ellipsoids. Moreover, the group-B chondrules show relatively small values of B/A ($\sim 0.75$), but ${\rm C/B} \simeq 1$. These shapes are called as ``prolate." Such not-oblate chondrules cannot be explained considering only the effect of the ram pressure. In order to produce the not-oblate molten droplet, we must investigate other effects which give some dynamical effect to the droplet (e.g., rotation of the droplet). In our forthcoming paper, we are planing to investigate the rapidly rotating droplets exposed to the gas flow. The dust rotation causes the three-dimensional deformation depending on the rotation axis. Therefore, the analysis based on the assumption of axis-symmetry would not be used. Since our numerical code has been developed on the concept of the three-dimension, any special improvements are not required for the application to the rapidly rotating droplets.

\subsection{Ram Pressure at Re-solidification?}
The precursor dust particles moving through the nebula gas are decelerated by the ram pressure of the gas flow. Therefore, the relative velocity $v_{\rm rel}$ between them becomes small with time. It indicates that the gas ram pressure $p_{\rm fm} = \rho_{\rm g} v_{\rm rel}^2$ at the time when the molten droplet re-solidifies would be smaller than that when it melts first. 

\begin{figure}[]
\center
\begin{tabular}{cc}
	\vspace{5mm}
	\begin{minipage}{.45\linewidth}
		\begin{center}(a) low dust-to-gas mass ratio\end{center}
		\vspace{0mm}
		\includegraphics[width=\linewidth, angle=-0.]{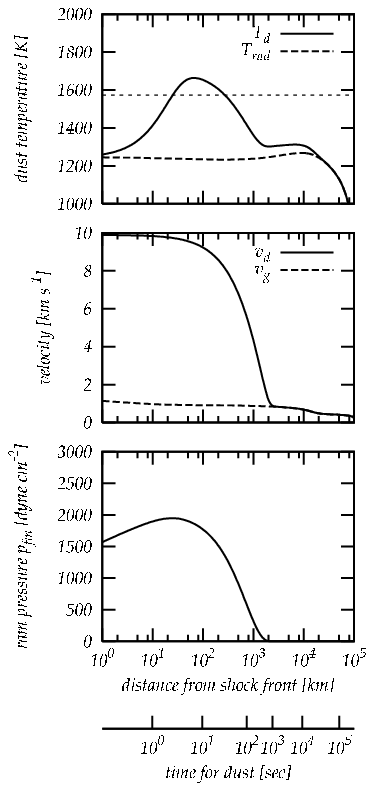}
	\end{minipage} &
	\begin{minipage}{.45\linewidth}
		\begin{center}(b) high dust-to-gas mass ratio\end{center}
		\vspace{0mm}
		\includegraphics[width=\linewidth, angle=-0.]{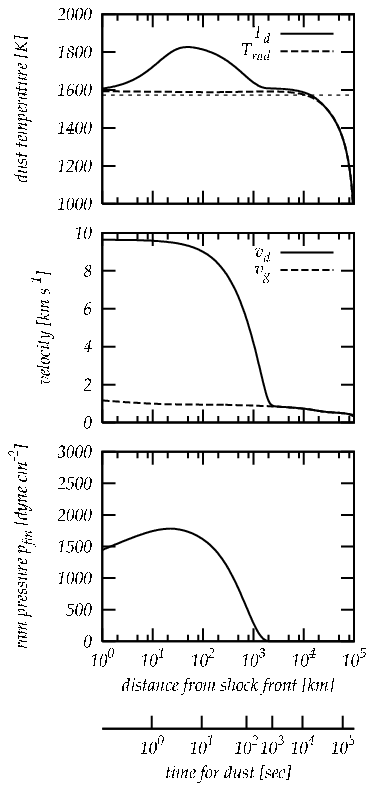}
	\end{minipage} \\
\end{tabular}
\caption{Dynamical/thermodynamical evolutions of the precursor dust particles in the post-shock region are plotted as a function of the distance from the shock front. Top: the dust temperature, $T_{\rm d}$, and the radiation temperature $T_{\rm rad}$. Middle: velocities of dust particle, $v_{\rm d}$, and gas $v_{\rm g}$. Bottom: the ram pressure $p_{\rm fm}$. The time for dust particle after passage of the shock front is also displayed.}
\label{fig:pfm}
\end{figure}

Figure \ref{fig:pfm} shows the dynamical/thermodynamical evolutions of precursor dust particles in the post-shock region (Miura \& Nakamoto 2006). The dust radius is $1000 \, {\rm \mu m}$. The horizontal axis is the distance from the shock front $x$ in logarithmic scale. The top panel shows the dust temperature $T_{\rm d}$ and the radiation temperature defined by $T_{\rm rad} \equiv (\pi {\cal J} / \sigma_{\rm SB})^{1/4}$, where ${\cal J}$ is the mean intensity of the radiation field mainly emitted from the dust particles and the shocked gas in the post-shock region, and $\sigma_{\rm SB}$ is the Stefan-Boltzmann constant. The middle panel shows the dust velocity $v_{\rm d}$ and the gas velocity $v_{\rm g}$ against the shock front. The bottom panel shows the ram pressure acting on the dust particles. The shock velocity $v_{\rm s} = 10 \, {\rm km \, s^{-1}}$, the pre-shock gas number density $n_0 = 10^{14} \, {\rm cm^{-3}}$, and the power-law dust size distribution are assumed. The mean intensity ${\cal J}$ also depends on the dust-to-gas mass ratio $C_{\rm d}$. The left column indicates a case of the lower dust-to-gas mass ratio ($C_{\rm d} = 0.01$). The dust temperature increases up to about $1700 \, {\rm K}$ around $x \sim 10^2 \, {\rm km}$ and after that decreases. Although the relative velocity between the dust particle and the gas decreases monotonically with the distance, the ram pressure $p_{\rm fm}$ increases until $x \sim 10^2 \, {\rm km}$ because the gas density $\rho_{\rm g}$ increases, and after that decreases. However, it is found that the ram pressure does not decrease significantly when the dust temperature falls down to the melting point of silicates ($1573 \, {\rm K}$, thin dashed line, e.g., Tachibana \& Huss 2006). On the contrary, the right column is a more dusty case ($C_{\rm d} = 0.10$). In that case, the radiation temperature is stronger than the left panels as a result of the blanket effect and in this case, exceeds the melting point of silicates. Since the dust particles are also heated by the radiation field, they do not cool even if the relative velocity is almost zero. They can cool if the radiation field becomes weak. At that time, the ram pressure has not already taken place on the dust particles. To summarize, the ram pressure acting on the re-solidifing molten droplet can take various values depending on the shock conditions and dust models in the chondrule-forming region. Generally, the low dust-to-gas mass environment would result into the large ram pressure. 

\hspace{1cm}{\bf [Figure \ref{fig:pfm}]}

\subsection{Dust Particle Rotation\label{sec:dust_rotation}}
We investigated the hydrodynamics of molten droplet exposed to the gas flow. However, in the shock-wave heating model, the situation where the rapidly rotating droplets are exposed to the high-velocity gas flow can be considered. Some possibilities for the origin of the dust rotation can be considered, for example, the net torque when the droplet fragments by the gas ram pressure (Susa \& Nakamoto 2002), the interaction between the dust particles and the ambient gas flow, and the dust-dust collision. Therefore, we have to take into account the effect of the rotation. When the rotation axis is perpendicular to the gas flow, the three-dimensional effect is expected in the droplet shapes. Our code has been already developed in the concept of the three-dimensional calculations. We are planing to examine the case that the rotating droplets are exposed to the high-velocity gas flow in our forthcoming paper.

\section{Conclusions}
We have developed the non-linear, time-dependent, and three-dimensional hydrodynamic simulation code in order to investigate the hydrodynamics of molten droplets exposed to the high-velocity rarefied gas flow. Our numerical code is based on the concept of the CIP scheme, guarantees the exact mass conservation, and additionally, we introduced the anti-diffusion technique for suppressing the effect of the numerical diffusion at the discontinuity in density. We carried out the simulations for various droplet radii ($r_0 = 100 \, {\rm \mu m } - 2 \, {\rm cm}$). The other physical parameters were adopted typical values for the shock-wave heating condition (the momentum of molecular gas flow $p_{\rm fm} = 4000 \, {\rm dyne \, cm^{-2}}$) and the silicate melts (surface tension $\gamma = 400 \, {\rm dyne \, cm^{-1}}$, coefficient of viscosity $\mu_{\rm d} = 1.3 \, {\rm g \, cm^{-1} \, s^{-1}}$). We conclude as follows:
\begin{enumerate}
\item Since we considered that the gas ram pressure suddenly affects the initially spherical droplet, the droplets whose radii are smaller than $5000 \, {\rm \mu m}$ showed the vibrational motion (deformation by the ram pressure and recovery motion by the surface tension). The vibration gradually dissipates by the viscosity and the droplets tend to settle in the steady states. 
\item The degree of the deformation at the steady state (or time-averaged value) showed a good agreement with the linear solution derived by Sekiya et al. (2003) if the droplet radii are smaller than $1000 \, {\rm \mu m}$. For larger droplets, the final shapes are far from the linear solution because the non-linear term in the hydrodynamical equation dominates, which terms were ignored in the derivation of the linear solution. 
\item If the droplet radii are larger than $5000 \, {\rm \mu m}$, the hydrostatic pressure inside the eddies of the droplet internal flow becomes almost zero. In this region, the phase transition from liquid to vapor would occur and some bubbles would be generated in the droplet (cavitation). It is considered that the cavitation causes the fragmentation of the droplets. 
\item When the droplet radius is larger than 1 cm, the droplet fragments directly by the gas ram pressure. We found that the parent droplet breaks up to many smaller pieces. If these pieces collide again behind the parent droplet, the compound chondrules, which are thought to have collided two independent chondrules, might be formed. 
\item We considered that the fragmentation through the cavitation might be a new mechanism to determine the maximum size of chondrules. Using the order of magnitude estimation, we found that the cavitation can easily occur in the low-viscosity droplets. We discussed the possibility that this relation can explain why chondrules in some carbonaceous chondrites are smaller than that in other chondrites. 
\item We compared our simulation results of the molten droplets with chondrules measured by Tsuchiyama et al. (2003) in the external shape, and found that the variety of chondrule shapes cannot be re-produced only by the effect of the ram pressure of the gas flow. We pointed out the importance of other mechanisms, e.g., the rotation of the droplets. 
\end{enumerate}

\section*{Acknowledgments:}
We greatly appreciate Prof. Akira Tsuchiyama for his helpful comments and offering the observational data of chondrules. We are also grateful to Dr. Hidemi Mutsuda for many useful information about the CIP scheme described in his doctor thesis. HM was supported by the Research Fellowship of Japan Society for the Promotion of Science for Young Scientists. TN was partially supported by the Ministry of Education, Science, Sports and Culture, Grant-in-Aid for Scientific Research (C), 1754021.

\appendix
\section{Numerical Method in Hydrodynamics\label{appendix:numerical_method_hydro}}
In this appendix, we briefly explain the strategies of the numerical schemes that we adopted in our model. The hydrodynamical equations except for the equation of continuity were separated into two phases; the advection phase (\S \ref{appendix:advection_phase}) and the non-advection phase (\S \ref{appendix:non-advection_phase}). The equation of continuity was solved in the conservative form (\S \ref{appendix:r-cip_scheme}). We also describe the method to keep the incompressibility of the fluid in \S \ref{appendix:divergence_free} and the model parameters we adopted in this study in \S \ref{appendix:model_parameter}. 

\subsection{Advection Phase}
\label{appendix:advection_phase}

\subsubsection{CIP scheme}
\label{appendix:cip_scheme}
The CIP scheme is one of the high-accurate numerical schemes for solving the advection equation (Yabe et al. 2001). In one-dimension, the advection equation is written as
\begin{equation}
\frac{ \partial f }{ \partial t } + u \frac{ \partial f }{ \partial x } = 0, 
\label{eq:1D_advection_eq}
\end{equation}
where $f$ is a scaler variable of the fluid (e.g., density), $u$ is the fluid velocity toward the $x$-direction, and $t$ is the time. When the velocity $u$ is constant, the exact solution of Eq. (\ref{eq:1D_advection_eq}) is given by $f (x; t) = f (x - u t; 0)$, which indicates a simple translational motion of the spatial profile of $f$ with the velocity $u$. 

\begin{figure}[]
\center
\includegraphics[scale=1.0, angle=0.]{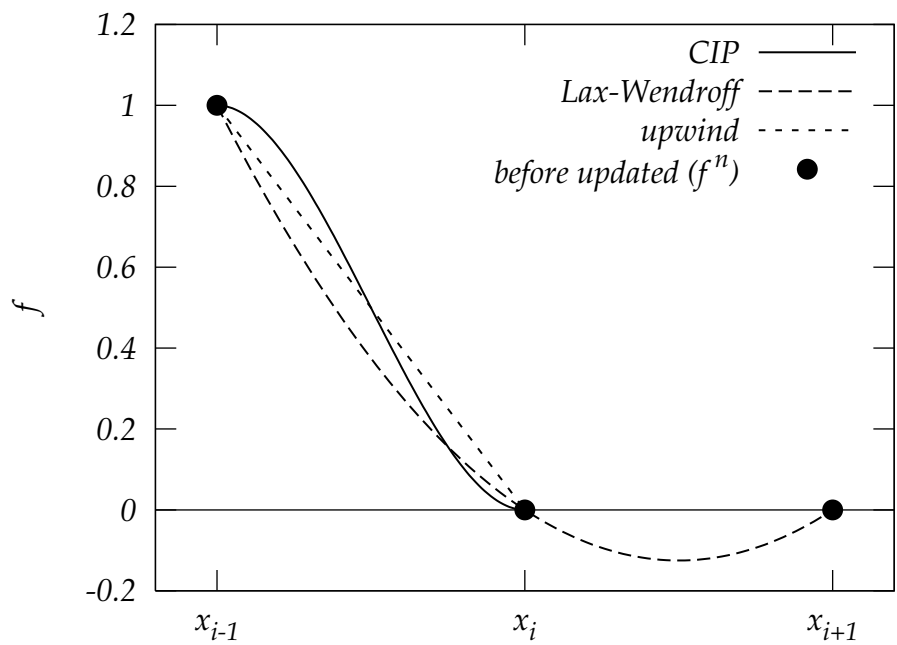}
\caption{Interpolate functions with various methods: CIP (solid), Lax-Wendroff (dashed), and first-order upwind (dotted). The filled circles indicate the values of $f$ defined on the digitized grid points $x_{i-1}$, $x_{i}$, and $x_{i+1}$ before updated.}
\label{fig:interpolate}
\end{figure}

Consider that at the time step $n$, the values of $f$ on the computational grid points $x_{i-1}$, $x_{i}$, and $x_{i+1}$ ($f_{i-1}^{n}$, $f_{i}^{n}$, and $f_{i+1}^{n}$) are given as the filled circles in Figure \ref{fig:interpolate} and the updated value at $x = x_{i}$ ($f_{i}^{n+1}$) is now required, where the updated value indicates the value at the next time step $n+1$. The time interval between these time steps is set to be $\Delta t$. From the solution of Eq. (\ref{eq:1D_advection_eq}), we can obtain $f_{i}^{n+1}$ by just calculating $f^{n}$ at the upstream point $x = x_{i} - u \Delta t$. If the upstream point locates between $x_{i-1}$ and $x_{i}$, we have to interpolate $f^{n}$ with an appropriate mathematical function composed of $f_{i-1}^{n}$, $f_{i}^{n}$, and so forth. There are some variations of the numerical solvers by the difference of the interpolate function $F_{i}(x)$. One of them is the first-order upwind scheme, which interpolates $f^{n}$ by a linear function and satisfies following two constraints; $F_{i}(x_{i-1}) = f_{i-1}^{n}$ and $F_{i}(x_{i}) = f_{i}^{n}$. The other variation is the Lax-Wendroff scheme, which uses a quadratic polynomial satisfying three constraints; $F_{i}(x_{i-1}) = f_{i-1}^{n}$, $F_{i}(x_{i}) = f_{i}^{n}$, and $F_{i}(x_{i+1}) = f_{i+1}^{n}$. 

On the contrary, the CIP scheme interpolates values using a cubic polynomial, which satisfies following four constraints; $F_{i}(x_{i-1}) = f_{i-1}^{n}$, $F_{i}(x_{i}) = f_{i}^{n}$, $\partial F_{i} / \partial x (x_{i-1}) = f_{x,i-1}^{n}$, and $\partial F_{i} / \partial x (x_{i}) = f_{x,i}^{n}$, where $f_{x} \equiv \partial f / \partial x$ is the spatial gradient of $f$. The way to update $f_{x}$ is as follows. Differentiating Eq. (\ref{eq:1D_advection_eq}), we obtain 
\begin{equation}
\frac{ \partial f_x }{ \partial t } + u \frac{ \partial f_x }{ \partial x } = - f_x \frac{ \partial u }{ \partial x },
\label{eq:1D_advection_eq_dif}
\end{equation}
where the second term of the left hand side indicates the advection and the right hand side indicates the non-advection term. The interpolate function for the advection of $f_x$ is given by $\partial F_i / \partial x$. The non-advection term can be solved analytically by considering that $\partial u / \partial x$ is constant. Additionally, there is an oscillation preventing scheme in the concept of the CIP scheme, in which the rational function is used as the interpolate function. The rational function is written as (Xiao et al. 1996a)
\begin{equation}
F_{i}(x) = \frac{ a_{i} (x - x_{i})^3 + b_{i} (x - x_{i})^2 + c_{i} (x - x_{i}) + f_{i}^{n} }{ 1 + \alpha_{i} \beta_{i} (x - x_{i}) }, 
\label{eq:rcip}
\end{equation}
where $a_{i}$, $b_{i}$, $c_{i}$, $\alpha_{i}$, and $\beta_{i}$ are the coefficients which are determined from $f_{i-1}^{n}$, $f_{i}^{n}$, $f_{x,i-1}^{n}$, and $f_{x,i}^{n}$. The expressions of these coefficients are shown in Xiao et al. (1996a). This scheme is called as the R-CIP scheme. 

Fig. \ref{fig:interpolate} also shows the interpolate functions with various methods; CIP (solid), Lax-Wendroff (dashed), and first-order upwind (dotted). In this figure, we assume $f_{x}^{n} = 0$ at all grid points for the CIP scheme. The difference in the numerical solutions of Eq. (\ref{eq:1D_advection_eq}) is discussed in \S \ref{appendix:test_calculation}.

\hspace{1cm}{\bf [Figure \ref{fig:interpolate}]}

\subsubsection{CIP-CSL2 scheme}
\label{appendix:r-cip_scheme}
\begin{figure}[]
\center
\includegraphics[scale=1., angle=0.]{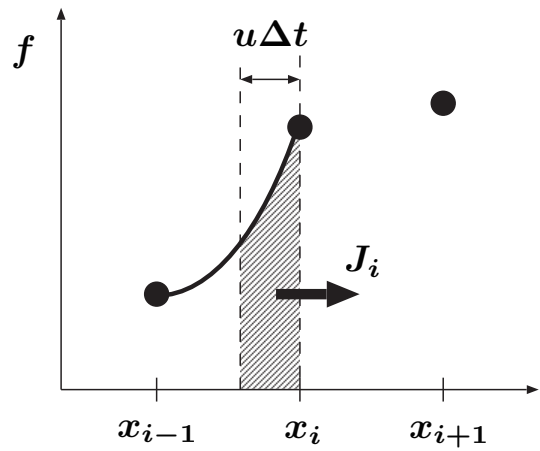}
\caption{The schematic picture of the CIP-CSL2 scheme.}
\label{fig:cip-csl2}
\end{figure}

The CIP-CSL2 scheme is one of the numerical schemes to solve the conservative equation. In one-dimension, the conservative equation is written as
\begin{equation}
\frac{\partial f}{\partial t} + \frac{\partial (u f)}{\partial x} = 0.
\label{eq:conservative_one-dimension}
\end{equation}
Integrating Eq. (\ref{eq:conservative_one-dimension}) over $x$ from $x_{i}$ to $x_{i+1}$, we obtain
\begin{equation}
\frac{\partial \sigma_{i+1/2}}{\partial t} + \Big[ u f \Big]_{x_{i}}^{x_{i+1}} = 0, 
\label{eq:conservative_one-dimension_int}
\end{equation}
where $\sigma_{i+1/2} \equiv \int_{x_{i}}^{x_{i+1}} f dx$. Since the physical meaning of $u f$ in the second term of the left hand side is the flux of $\sigma$ per unit area and per unit time, the time evolution of $\sigma$ is determined by
\begin{equation}
\sigma_{i+1/2}^{n+1} = \sigma_{i+1/2}^{n} - J_{i+1} + J_{i}, 
\label{eq:conservative_one-dimension_evo}
\end{equation}
where $J_{i} \equiv \int_{t^{n}}^{t^{n+1}} u f dt$ is the transported value of $\sigma$ from $x < x_{i}$ to $x > x_{i}$ within $\Delta t$, and it corresponds to the grayed region in Figure \ref{fig:cip-csl2}, where the filled circles indicate $f^{n}$ and the thick solid curve is the interpolate function $F_{i}(x)$ for a region $x_{i-1} < x < x_{i}$. We assume $u_{i} > 0$ in this figure. 

The CIP-CSL2 scheme uses the integrated function $D_{i}(x) \equiv \int_{x_{i-1}}^{x} F_{i} (x) dx$ for the interpolation when $u_{i} > 0$. The function $D_{i}(x)$ is a cubic polynomial satisfying following four constraints; $D_{i}(x_{i-1}) = 0$, $D_{i}(x_{i}) = \sigma_{i-1/2}$, $\partial D_{i} / \partial x (x_{i-1}) = F_{i} (x_{i-1}) = f_{i-1}$, and $\partial D_{i} / \partial x (x_{i}) = F_{i}(x_{i}) = f_{i}$. Moreover, since Eq. (\ref{eq:conservative_one-dimension}) can be rewritten as the same form of Eq. (\ref{eq:1D_advection_eq_dif}), we can obtain the updated value $f^{n+1}$ as well as $f_{x}^{n+1}$ in the CIP scheme (\S \ref{appendix:cip_scheme}). Additionally, there is an oscillation preventing scheme in the concept of the CIP-CSL2 scheme, in which the rational function is used for the function $D_{i}(x)$. The rational function is written as (Nakamura et al. 2001)
\begin{equation}
D_{i}(x) = \frac{ a_{i} (x - x_{i})^3 + b_{i} (x - x_{i})^2 + c_{i} (x - x_{i}) }{ 1 + \alpha_{i} \beta_{i} (x - x_{i}) }, 
\label{eq:rcip}
\end{equation}
where $a_{i}$, $b_{i}$, $c_{i}$, $\alpha_{i}$, and $\beta_{i}$ are the coefficients which are determined from $f_{i-1}$, $f_{i}$, and $\sigma_{i-1/2}$. The expressions of these coefficients are shown in Nakamura et al. (2001). This scheme is called as the R-CIP-CSL2 scheme. The example of the test calculation is shown in \S \ref{appendix:test_calculation}. 


\hspace{1cm}{\bf [Figure \ref{fig:cip-csl2}]}

\subsubsection{Anti-Diffusion}
\label{appendix:anti-diffusion}
In order to keep the sharp discontinuity in the profile of $\phi$, we explicitly add the diffusion term with a negative diffusion coefficient $\alpha$ to the CIP-CSL2 scheme. In our model, we have an additional diffusion equation about $\sigma$ (see Eq. \ref{eq:conservative_one-dimension_int}) as
\begin{equation}
\frac{\partial \sigma}{\partial t} = \frac{\partial}{\partial x} \bigg( \alpha \frac{\partial \sigma}{\partial x} \bigg).
\label{eq:anti-diffusion}
\end{equation}
Eq. (\ref{eq:anti-diffusion}) can be separated into two equations as
\begin{equation}
\frac{ \partial \sigma}{ \partial t } = - \frac{ \partial J' }{ \partial x }, 
\label{eq:anti-diffusion_a}
\end{equation}
\begin{equation}
J' = - \alpha \frac{ \partial \sigma }{ \partial x }, 
\label{eq:anti-diffusion_b}
\end{equation}
where $J'$ indicates the flux per unit area and per unit time. Using the finite difference method, we obtain
\begin{equation}
\sigma_{i+1/2}^{**} = \sigma_{i+1/2}^{*} - ( \hat{J}'_{i+1} - \hat{J}'_{i} ) , 
\label{eq:anti-diffusion_a_diff}
\end{equation}
\begin{equation}
\hat{J}'_{i} = - \hat{\alpha}_{i} (\sigma_{i+1/2} - \sigma_{i-1/2}), 
\label{eq:anti-diffusion_b2_diff}
\end{equation}
where $\hat{J}' \equiv J' / (\Delta x / \Delta t)$ is the mass flux which has the same dimension of $\sigma$ and $\hat{\alpha} \equiv \alpha / (\Delta x^2 / \Delta t)$ is the dimensionless diffusion coefficient. The superscript $*$ and $**$ indicate the time step just before and after the anti-diffusion. However, if we use Eq. (\ref{eq:anti-diffusion_b2_diff}) directly, we cannot obtain appropriate solutions. Therefore, we make some inventions in Eq. (\ref{eq:anti-diffusion_b2_diff}) to obtain the mass flux $\hat{J}'$. 

In this paper, we calculate $\hat{J}'$ as
\begin{equation}
\hat{J}'_{i} = - \hat{\alpha}_{i} \times {\rm minmod} ( S_{i-1} , S_{i} , S_{i+1} ), 
\label{eq:anti-diffusion_mass-flux}
\end{equation}
where $S_{i} \equiv \sigma_{i+1/2} - \sigma_{i-1/2}$. The minimum modulus function (minmod) is often used in the concept of the flux limiter and has a non-zero value of ${\rm sign}(a) \, {\rm min} ( |a|, |b|, |c| )$ only when $a$, $b$, and $c$ have same sign. The value of the diffusion coefficient $\hat{\alpha}$ is also important. Basically, we take $\hat{\alpha} = -0.1$ for the anti-diffusion. Here, it should be noted that $\sigma$ takes the limited value as $0 \le \sigma \le \sigma_{\rm m}$, where $\sigma_{\rm m}$ is the initial value for inside of the droplet. The undershoot ($\sigma < 0$) or overshoot ($\sigma > \sigma_{\rm m}$) are physically incorrect solutions. To avoid that, we replace $\hat{\alpha}_{i} = 0.1$ only when $\sigma_{i-1/2}$ or $\sigma_{i+1/2}$ are out of the appropriate range. 

We insert the anti-diffusion phase after the CIP-CSL2 scheme is completed. We also have the anti-diffusion for other directions ($y$ and $z$) in the same manner. The test calculations for one-dimensional conservative equation with and without the anti-diffusion are shown in \S \ref{appendix:test_calculation}.

\subsubsection{Test Calculation}
\label{appendix:test_calculation}
\begin{figure}[]
\center
\begin{tabular}{cc}
	\vspace{5mm}
	\begin{minipage}{.45\linewidth}
		\begin{center}(a) first-order upwind\end{center}
		\vspace{-3mm}
		\includegraphics[width=\linewidth, angle=-0.]{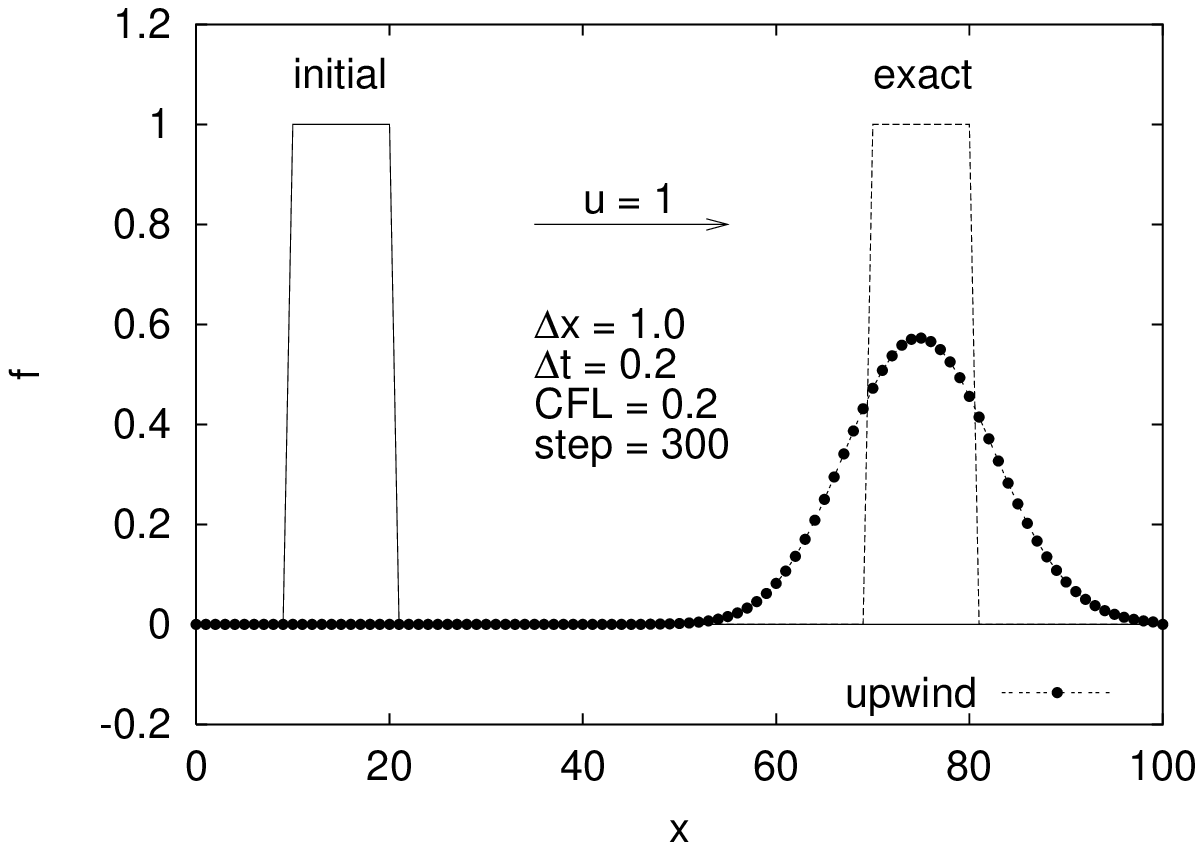}
	\end{minipage} &
	\begin{minipage}{.45\linewidth}
		\begin{center}(b) Lax-Wendroff\end{center}
		\vspace{-3mm}
		\includegraphics[width=\linewidth, angle=-0.]{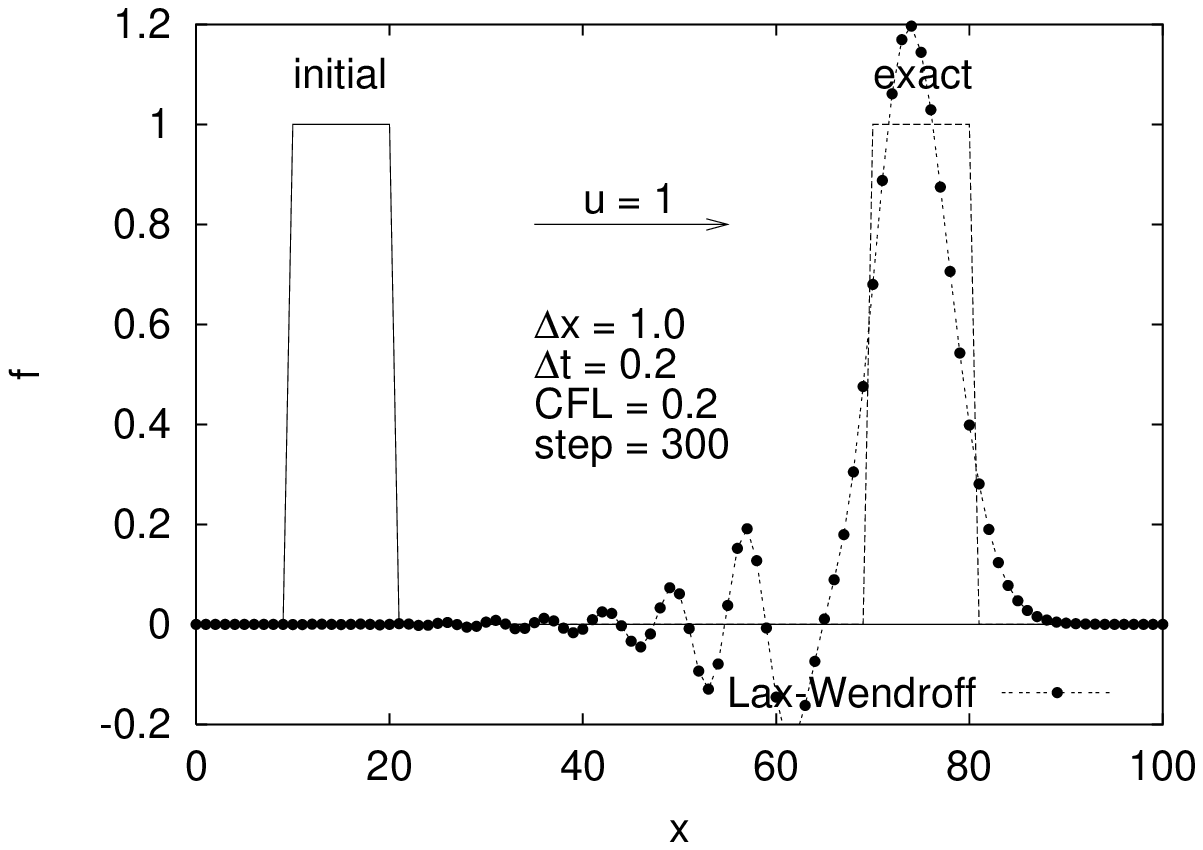}
	\end{minipage} \\
	\vspace{5mm}
	\begin{minipage}{.45\linewidth}
		\begin{center}(c) CIP\end{center}
		\vspace{-3mm}
		\includegraphics[width=\linewidth, angle=-0.]{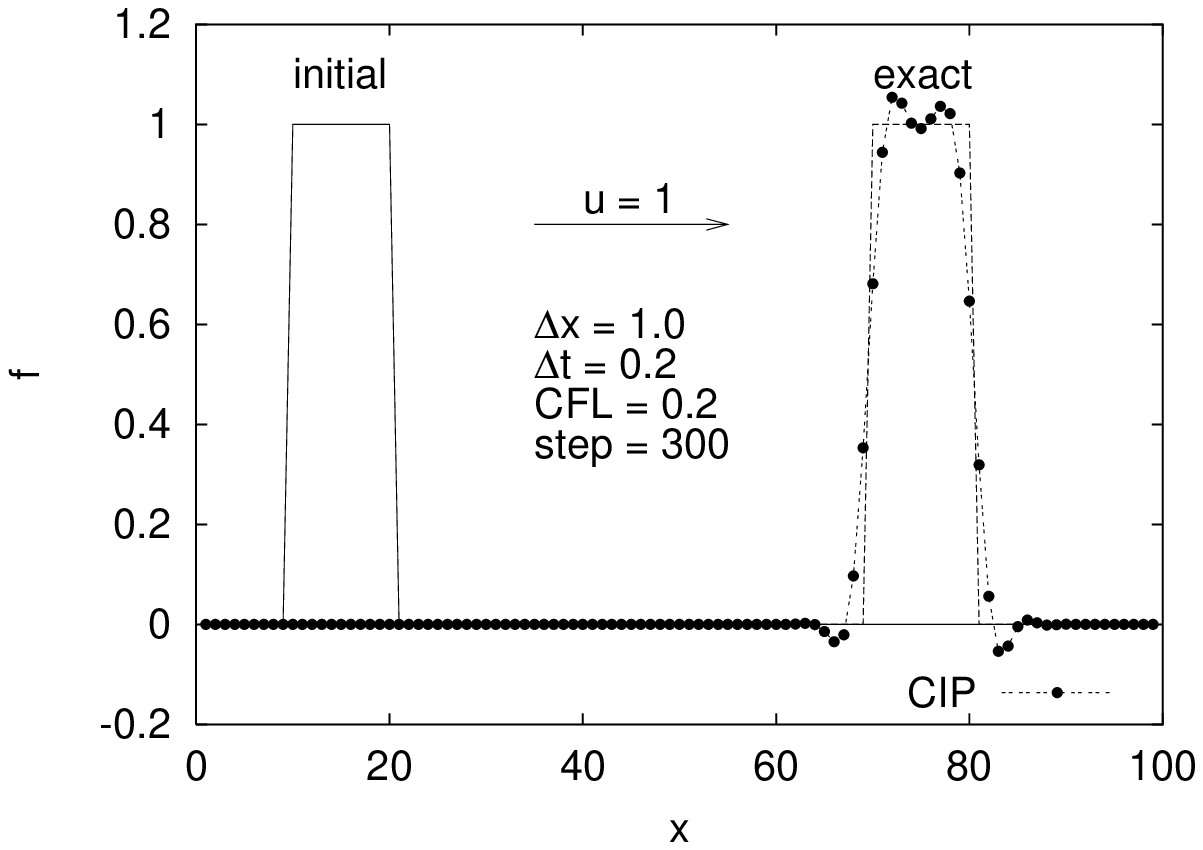}
	\end{minipage} &
	\begin{minipage}{.45\linewidth}
		\begin{center}(d) R-CIP\end{center}
		\vspace{-3mm}
		\includegraphics[width=\linewidth, angle=-0.]{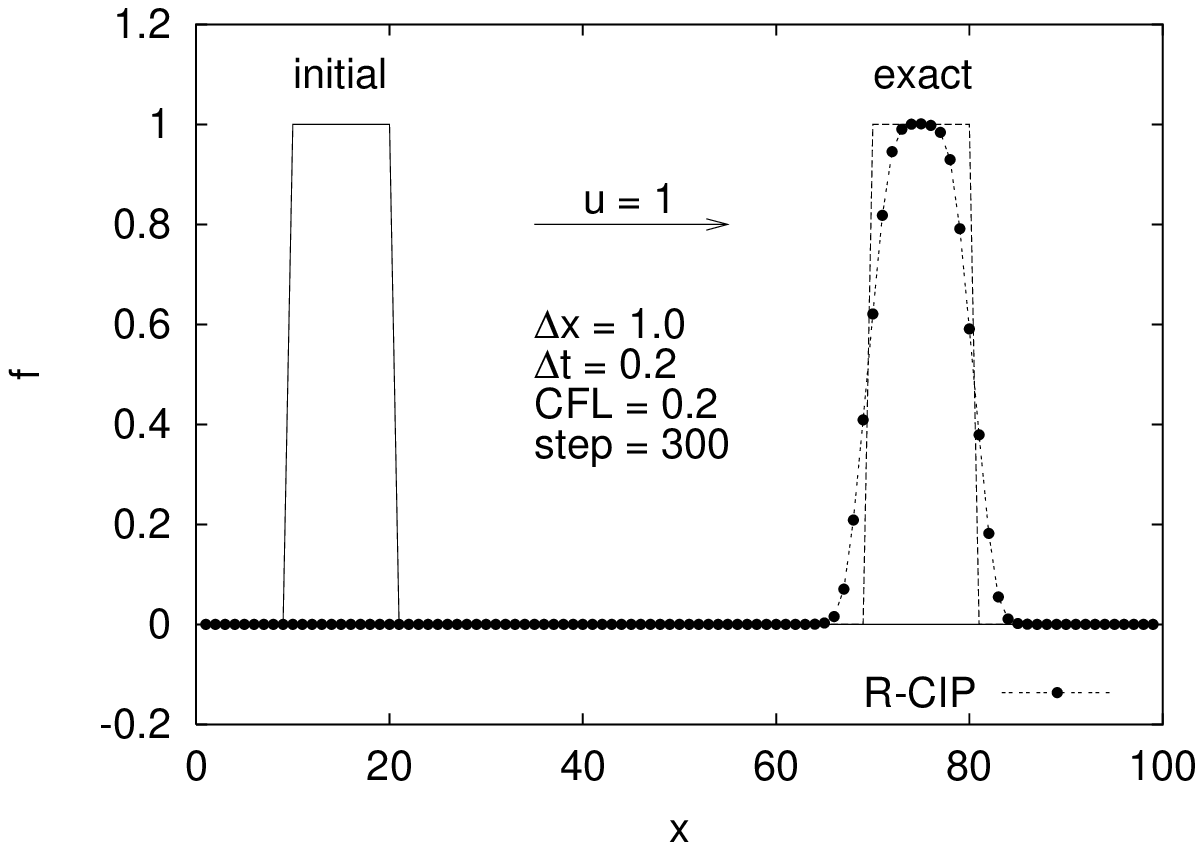}
	\end{minipage} \\
	\vspace{5mm}
	\begin{minipage}{.45\linewidth}
		\begin{center}(e) R-CIP-CSL2\end{center}
		\vspace{-3mm}
		\includegraphics[width=\linewidth, angle=-0.]{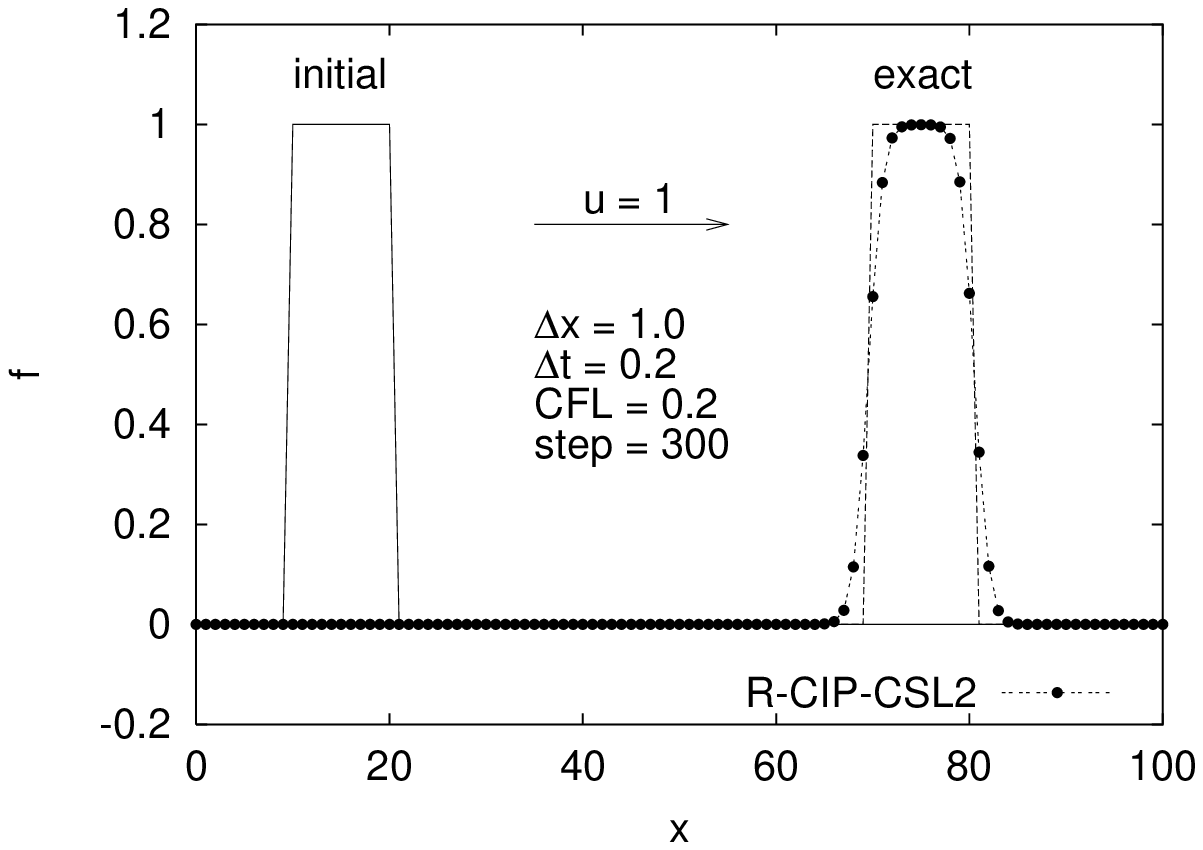}
	\end{minipage} &
	\begin{minipage}{.45\linewidth}
		\begin{center}(f) R-CIP-CSL2 + anti-diffusion\end{center}
		\vspace{-3mm}
		\includegraphics[width=\linewidth, angle=-0.]{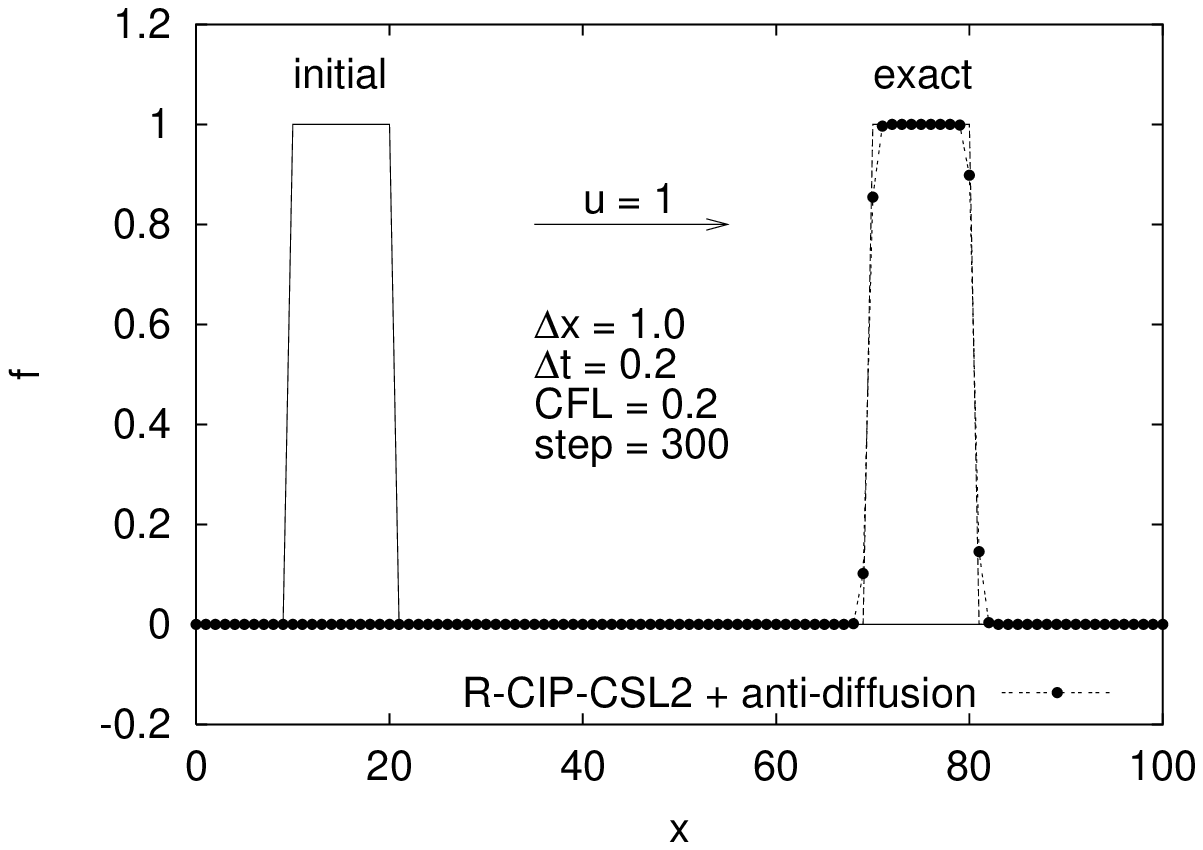}
	\end{minipage} \\
\end{tabular}
\caption{Numerical solutions of the one-dimensional advection or conservative equation solved by various schemes: (a) first-order upwind, (b) Lax-Wendroff, (c) CIP, (d) R-CIP, (e) R-CIP-CSL2 without anti-diffusion, and (f) R-CIP-CSL2 with anti-diffusion.}
\label{fig:advection_solution}
\end{figure}

In order to demonstrate the advantage of the CIP scheme, we carried out one-dimensional advection calculations with various numerical schemes. Figure \ref{fig:advection_solution} shows the spatial profiles of $f$ of the test calculations. The horizontal axis is the spatial coordinate $x$. The initial profile is given by the thin solid line, which indicates a rectangle wave. We set the fluid velocity $u = 1$, the intervals of the grid points $\Delta x = 1$, and the time step for the calculation $\Delta t = 0.2$. These conditions give the CFL number $\nu \equiv u \Delta t / \Delta x = 0.2$, which indicates that the profile of $f$ moves 0.2 times the grid interval per time step. Therefore, the right side of the rectangle wave will reach $x = 80$ after 300 time steps and the dashed line indicates the exact solution. The filled circles indicate the numerical results after 300 time steps. 

The upwind scheme does not keep the rectangle shape and the profile becomes smooth by the numerical diffusion (panel a). In the Lax-Wendroff scheme, the numerical oscillation appears behind the real wave (panel b). Comparing with above two schemes, the CIP scheme seems to show better solution, however, some undershoots ($f < 0$) or overshoots ($f > 1$) are observed in the numerical result (panel c). In the R-CIP scheme, although the faint numerical diffusion has still remained, we obtain the excellent solution comparing with the exact solution (panel d). 

We also show the numerical results of the one-dimensional conservative equation (Eq. \ref{eq:conservative_one-dimension}). We use the same conditions with the one-dimensional advection equation (Eq. \ref{eq:1D_advection_eq}). Note that Eq. (\ref{eq:conservative_one-dimension}) corresponds to Eq. (\ref{eq:1D_advection_eq}) when the velocity $u$ is constant. The panel (e) shows the results of the R-CIP-CSL2 scheme, which is similar to that of the R-CIP scheme. In the panel (f), we found that the combination of the R-CIP-CSL2 scheme and the anti-diffusion technique (\S \ref{appendix:anti-diffusion}) shows the excellent solution in which the numerical diffusion is prevented effectively.

\hspace{1cm}{\bf [Figure \ref{fig:advection_solution}]}

\subsection{Non-Advection Phase}
\label{appendix:non-advection_phase}

\subsubsection{C-CUP scheme}
\label{appendix:c-cup_scheme}
Using the finite difference method to Eq. (\ref{eq:non-advection_phase}), we obtain
\begin{equation}
\frac{ \mbox{\boldmath $u$}^{**} - \mbox{\boldmath $u$}^{*} }{ \Delta t } = - \frac{ \mbox{\boldmath $\nabla$} p^{**} }{ \rho^{*} } + \frac{ \mbox{\boldmath $Q$} }{ \rho^{*} }, ~~~
\frac{ p^{**} - p^{*} }{ \Delta t } = - \rho^{*} c_{\rm s}^2 \mbox{\boldmath $\nabla$} \cdot \mbox{\boldmath $u$}^{**}, 
\label{eq:pressure_based_algorithm}
\end{equation}
where the superscripts $*$ and $**$ indicate the times before and after calculating the non-advection phase, respectively. Since the sound speed can be very large in the incompressible fluid, the term related to the pressure should be solved implicitly. In order to obtain the implicit equation for $p^{**}$, we take the divergence of the left equation and substitute $\mbox{\boldmath $u$}^{**}$ into the right equation. Then we obtain an equation
\begin{equation}
\mbox{\boldmath $\nabla$} \cdot \Bigg( \frac{ \mbox{\boldmath $\nabla$} p^{**} }{ \rho^{*} } \Bigg) = \frac{ p^{**} - p^{*} }{ \rho^{*} c_{\rm s}^2 \Delta t^2 } + \frac{ \mbox{\boldmath $\nabla$} \cdot \mbox{\boldmath $u$}^{*} }{ \Delta t } + \mbox{\boldmath $\nabla$} \cdot \Bigg( \frac{ \mbox{\boldmath $Q$} }{ \rho^{*} } \Bigg). 
\label{eq:c-cup_scheme}
\end{equation}
The problem to solve Eq. (\ref{eq:c-cup_scheme}) resolves itself into to solve a set of linear algebraic equations in which the coefficients become an asymmetric sparse matrix. After $p^{**}$ is solved, we can calculate $\mbox{\boldmath $u$}^{**}$ by solving the left equation in Eq. (\ref{eq:pressure_based_algorithm})\footnote{In our model, we neglect the viscous term in $\mbox{\boldmath $Q$}$ when calculating Eq. \ref{eq:c-cup_scheme}. In the original C-CUP scheme developed by Yabe \& Wang (1991), the other terms (correspond to the surface tension and the ram pressure, in our model) are also ignored when calculating the pressure.}.

\subsubsection{Ram Pressure of Gas Flow}
\label{appendix:gas_drag_force}
Consider that the molecular gas flows for the positive direction of the $x$-axis. The $x$-component of the ram pressure $F_{{\rm g},x}$ is given by
\begin{equation}
F_{{\rm g},x} = p_{\rm fm} \delta (x - x_{\rm i}), 
\label{eq:gas_drag_force_x}
\end{equation}
where $x_{\rm i}$ is the position of the droplet surface. This equation can be separated into two equations as
\begin{equation}
\frac{ \partial M }{ \partial x } = - p_{\rm fm} \delta (x - x_{\rm i}), 
\label{eq:momentum_gas_flow}
\end{equation}
\begin{equation}
F_{{\rm g},x} = - \frac{ \partial M }{ \partial x }, 
\label{eq:gas_drag_force_x2}
\end{equation}
where $M$ is the momentum flux of the molecular gas flow. Eq. (\ref{eq:momentum_gas_flow}) means that the momentum flux terminates at the droplet surface. Eq. (\ref{eq:gas_drag_force_x2}) means that the decrease of the momentum flux per unit length corresponds to the ram pressure per unit area. 

Using the finite difference method to Eq. (\ref{eq:momentum_gas_flow}), we obtain
\begin{equation}
M_{i+1} = M_{i} - p_{\rm fm} ( \bar{\phi}_{i+1} - \bar{\phi}_{i} )
~~~ {\rm for} ~ \bar{\phi}_{i+1} \ge \bar{\phi}_{i}, 
\label{eq:momentum_gas_flow_diff}
\end{equation}
where $\bar{\phi}$ is the smoothed profile of $\phi$ (see appendix \ref{appendix:smoothing}), and $M_{i+1} = M_{i}$ for $\bar{\phi}_{i+1} < \bar{\phi}_{i}$ because the momentum flux does not increase when the molecular flow goes outward from inside of the droplet. Similarly, we obtain
\begin{equation}
F_{{\rm g},xi} = - \frac{ M_{i} - M_{i-1} }{ \Delta x }
\label{eq:gas_drag_force_x2_diff}
\end{equation}
from Eq. (\ref{eq:gas_drag_force_x2}). The momentum flux at upstream is $M_0 = p_{\rm fm}$. First, we solve Eq. (\ref{eq:momentum_gas_flow_diff}) and obtain the spatial distribution of the molecular gas flow in all computational domain. After that, we calculate the ram pressure by Eq. (\ref{eq:gas_drag_force_x2_diff}).

\subsubsection{Surface Tension\label{appendix:surface_tension}}
The surface tension is the normal force per unit interfacial area. Brackbill et al. (1992) introduced a method to treat the surface tension as a volume force by re-placing the discontinuous interface to the transition region which has some width. According to them, the surface tension is expressed as
\begin{equation}
\mbox{\boldmath $F_{\rm s}$} = \gamma \kappa \mbox{\boldmath $\nabla$} \phi / [ \phi ] ,
\label{eq:surface_tension_brackbill}
\end{equation}
where $[ \phi ]$ is the jump in color function at the interface between the droplet and the ambient gas (in our definition, we obtain $[ \phi ] = 1$). The curvature is given by
\begin{equation}
\kappa = - ( \mbox{\boldmath $\nabla$} \cdot \hat{\mbox{\boldmath $n$}} ), 
\label{eq:curvature_brackbill}
\end{equation}
where
\begin{equation}
\hat{\mbox{\boldmath $n$}} = \mbox{\boldmath $\nabla$} \phi / | \mbox{\boldmath $\nabla$} \phi |. 
\label{eq:n_hat_brackbill}
\end{equation}
The finite difference method of Eq. (\ref{eq:n_hat_brackbill}) is shown in Brackbill et al. (1992) in detail. When we calculate the surface tension, we use the smoothed profile of $\phi$ (see appendix \ref{appendix:smoothing}).

\subsubsection{Smoothing\label{appendix:smoothing}}
We can obtain the numerical results keeping the sharp interface between the droplet and the ambient region. However, the smooth interface is suitable for calculating the smooth surface tension. We use the smoothed profile of $\phi$ only at the time to calculate the surface tension and the ram pressure acting on the droplet surface. In this study, the smoothed color function $\bar{\phi}$ is calculated by
\begin{equation}
\bar{\phi} = \frac{1}{2} \phi_{i,j,k} + \frac{1}{2} \frac{ \phi_{i,j,k} + C_1 \sum_{L_1}^6 \phi_{L_1} + C_2 \sum_{L_2}^{12} \phi_{L_2} + C_3 \sum_{L_3}^8 \phi_{L_3} }{ 1 + 6 C_1 + 12 C_2 + 8 C_3 }, 
\label{eq:smoothing}
\end{equation}
where $L_1$, $L_2$, and $L_3$ indicate grid indexes of the nearest, second nearest, and third nearest from the grid point $(i,j,k)$, for example, $L_1 = (i+1,j,k)$, $L_2 = (i+1,j+1,k)$, $L_3 = (i+1,j+1,k+1)$, and so forth. It is easily found that in the three-dimensional Cartesian coordinate system, there are six for $L_1$, twelve for $L_2$, and eight for $L_3$, respectively. The coefficients are set as
\begin{equation}
C_1 = 1 / ( 6 + 12 / \sqrt{2} + 8 / \sqrt{3} ), 
~~~
C_2 = C_1 / \sqrt{2}, 
~~~
C_3 = C_1 / \sqrt{3}.
\label{eq:smoothing_coefficients}
\end{equation}
We iterate the smoothing five times. Then, we obtain the smooth transition region of about twice grid interval width. We use the smooth profile of $\phi$ only when calculating the surface tension and the ram pressure. It should be noted that the original profile $\phi$ with the sharp interface is kept unchanged.

\subsection{Pressure Correction for Incompressible Fluid}
\label{appendix:divergence_free}
The molten chondrule precursor dust particles can be regarded as incompressible because of the large sound velocity (see \S \ref{sec:physical_model}). However, in the numerical solutions, it can occur that the velocity in the droplet calculated by Eq. (\ref{eq:pressure_based_algorithm}) has some divergence. In order to vanish it, we need to perform an extra iteration step, for example, as proposed by Chorin (1968). 

Although we obtained the velocity $\mbox{\boldmath $u$}^{**}$ after calculating Eq. (\ref{eq:pressure_based_algorithm}), the divergence of $\mbox{\boldmath $u$}^{**}$ inside the droplet is not small enough to be expected from its large sound speed (incompressibility). In order to obtain the proper velocity, we need to correct the pressure by some method and re-calculate the velocity by using the corrected pressure. Here, we write the velocity and pressure after the correction as $\mbox{\boldmath $\tilde{u}$}^{**}$ and $\tilde{p}^{**}$. Additionally, we define the correction of the pressure as $\delta p \equiv \tilde{p}^{**} - p^{**}$. The velocity should be changed according to the equation of motion. On the other hand, the pressure should be changed according to the equation of state. Therefore, we have the velocity/pressure correction method as (c.f., Eq. \ref{eq:pressure_based_algorithm})
\begin{equation}
\frac{ \mbox{\boldmath $\tilde{u}$}^{**} - \mbox{\boldmath $u$}^{**} }{ \Delta t } = - \frac{ \mbox{\boldmath $\nabla$} \delta p }{ \rho^{*} } , ~~~
\frac{ \tilde{p}^{**} - p^{**} }{ \Delta t } = - \rho^{*} c_{\rm s}^2 \mbox{\boldmath $\nabla$} \cdot \mbox{\boldmath $\tilde{u}$}^{**}.
\label{eq:divergence_free}
\end{equation}
We iterate these equations until $\tilde{p}^{**}$ converges. The same numerical scheme as applied for Eq. (\ref{eq:pressure_based_algorithm}) can be used to solve Eq. (\ref{eq:divergence_free}).


\subsection{Model Parameters}
\label{appendix:model_parameter}
In Table \ref{table:input_parameters}, parameters marked by (*) have no physical meanings. These values are empirical settings for the numerical simulation.

While we set the density of the ambient region $\rho_{\rm a} = 10^{-6} \, {\rm g \, cm^{-3}}$, the typical gas density of the protoplanetary disk is $\sim 10^{-9} \, {\rm g \, cm^{-3}}$ in the minimum mass solar nebula (Hayashi et al. 1985). There is a difference of about four orders of magnitude in the densities, however, the numerical results do not depend significantly on the ambient density because it is too small to affect the dynamics of the molten droplet. The stronger the density contrast between the droplet and the ambient region is, the more difficult the numerical simulation becomes. From above reasons, we adopted the relatively large value for the ambient density in the numerical simulation. 

While we set the sound speed of the ambient region $c_{\rm s,a} = 10^{-5} \, {\rm cm \, s^{-1}}$, the typical value of the disk gas is $c_{\rm s, disk} \simeq (k_{\rm B} T / m_{\rm H_2})^{1/2} \simeq 10^{5} \, {\rm cm \, s^{-1}}$, where $k_{\rm B}$ is the Stefan-Boltzmann constant, $T$ is the gas temperature (we substitute $T = 300 \, {\rm K}$), and $m_{\rm H_2}$ is the molecular weight of H$_2$. Remember that in this study, the disk gas around the molten droplet behaves like as a free molecular flow and does not follow the hydrodynamical equations (\S \ref{sec:physical_model}). Therefore, we do not need to consider the sound wave and the pressure change by the compression/expansion in the ambient region. We thought that the extremely low sound speed for the ambient region is one of the appropriate ways to express above features in the concept of the multi-fluids analysis. 

Finally, we set the viscous coefficient of the ambient region $\mu_{\rm a} = 10^{-2} \, {\rm g \, cm^{-1} \, s^{-1}}$ because it should not be zero for the stability of the calculations.

\section{Moment of Inertia for Linear Solution\label{appendix:moment_of_inertia}}
The external shapes obtained by our numerical simulations were approximated as three-axial ellipsoids using the moments of inertia. We compared our results with the linear solution derived by Sekiya et al. (2003) in this paper. When we obtain the axial ratio of the linear solution, the moment of inertia for the linear solution is needed. It can be calculated using the formalism by Sekiya et al. (2003) as follows. 

Sekiya et al. (2003) gave the distance between the dust surface to the origin of the spherical coordinates\footnote{Sekiya et al. (2003) chose the origin of the spherical coordinates as $dr_{\rm s} / d\theta = 0$ at $\theta = \pi / 2$.} as $r_{\rm s} (\theta) = r_0 ( 1 + W_e x_{\rm s} (\theta) )$, where $W_e = p_{\rm fm} r_0 / \gamma$ is the Weber number, $\theta$ is the angle from the radius of droplet to the negative $x$-direction in our coordinate, and $x_{\rm s}$ is the dimensionless parameter which was given as a function of $\theta$. The expression of $x_{\rm s}$ is given in Sekiya et al. (2003). The moment of inertia around the $x$-axis is calculated as
\begin{eqnarray}
I_{xx} &=& \rho_{\rm d} \int_0^{r_{\rm s} (\theta)} dr \int_0^{\pi} d\theta \int_0^{2 \pi} d\psi (r \sin \theta)^2 r^2 \sin \theta, \nonumber \\
&=& \frac{ 2 }{ 5 } \pi \rho_{\rm d} r_0^5 \sum_{m=0}^{5} {}_5 {\rm P}_m b_m W_e^m, 
\label{eq:moment_of_inertia_x}
\end{eqnarray}
where P is the permutation defined by ${}_n {\rm P}_r \equiv n! / (n-r)!$ and 
\begin{equation}
b_m \equiv \int_0^{\pi} x_{\rm s}^m (\theta) \sin^3 \theta d\theta. 
\label{eq:moment_of_inertia_x_coefficient}
\end{equation}
We obtain $b_0 = 4/3$. It can be easily confirmed that $I_{xx}$ has the same value of a sphere ($I = (8/15) \pi \rho_{\rm d} r_0^5$) when $W_e = 0$. For other coefficients, we calculate using an appropriate numerical integration method and obtain $b_1 = 0.42222\times10^{-1}$, $b_2 = 0.51562 \times 10^{-2}$, $b_3 = 0.74446 \times 10^{-4}$, $b_4 = 0.44106 \times 10^{-4}$, and $b_5 = -0.28485 \times 10^{-5}$.

\clearpage

\def\baselinestretch{1}\large\normalsize

\begin{table}[]
\caption{Input physical parameters for simulations of molten droplet exposed to the gas flow. The meanings of the parameters marked by (*) are shown in Appendix \ref{appendix:model_parameter}. }
\begin{center}
\begin{tabular}{lll}
\hline
parameter & sign & value \\
\hline \hline
momentum of gas flow & $p_{\rm fm}$ & $4000 \, {\rm dyne \, cm^{-2}}$ \\
surface tension & $\gamma$ & $400 \, {\rm dyne \, cm^{-1}}$ \\
viscosity of droplet & $\mu_{\rm d}$ & $1.3 \, {\rm g \, cm^{-1} \, s^{-1}}$   \\
density of droplet & $\rho_{\rm d}$ & $3 \, {\rm g \, cm^{-3}}$ \\
sound speed of droplet & $c_{\rm s,d}$ & $2 \times 10^{5} \, {\rm cm \, s^{-1}}$ \\
(*) density of ambient & $\rho_{\rm a}$ & $10^{-6} \, {\rm g \, cm^{-3}}$ \\
(*) sound speed of ambient & $c_{\rm s,a}$ & $10^{-5} \, {\rm cm \, s^{-1}}$ \\
(*) viscosity of ambient & $\mu_{\rm a}$ & $10^{-2} \, {\rm g \, cm^{-1} \, s^{-1}}$ \\
\hline
\end{tabular}
\end{center}
\label{table:input_parameters}
\end{table}

\clearpage

\clearpage

\clearpage

\clearpage

\clearpage

\clearpage

\clearpage

\clearpage

\clearpage

\clearpage

\clearpage

\clearpage

\clearpage

\end{document}